# A review of the effects of chemical and phase segregation on the mechanical behaviour of multi-phase steels

**Authors:**

Bernard Ennis[a, b*],

July 2014

a. *Tata Steel Research and Development, 1970 CA IJmuiden, The Netherlands*
b. *The School of Materials, University of Manchester, Oxford Road, Manchester, M13 9PL, UK*

*\* Corresponding author: bernard.ennis@tatasteel.com*

**Abstract**

In the drive towards higher strength alloys, a diverse range of alloying elements is employed to enhance their strength and ductility. Limited solid solubility of these elements in steel leads to segregation during casting which affects the entire down-stream processing and eventually the mechanical properties of the finished product. Although it is thought that the presence of continuous bands lead to premature failure, it has not been possible to verify this link. This poses as increasingly greater risk for higher alloyed, higher strength steels which are prone to centre-line segregation: it is thus vital to be able to predict the mechanical behaviour of multi-phase (MP) steels under loading.

This review covers the microstructure and properties of galvanised advanced high strength steels with particular emphasis to their use in automotive applications. In order to understand the origins of banding, the origins of segregation of alloying elements during casting and partitioning in the solid state will be discussed along with the effects on the mechanical behaviour and damage evolution under (tensile) loading. Attention will also be paid to the application of microstructural models in tailoring the production process to enable suppression of the effects of segregation upon banding. Finally, the theory and application of the experimental techniques used in this work to elucidate the structure and properties will be examined.

# Contents                                                                    Page

# List of figures

## List of tables

# Abbreviations used in this report

| Abbreviation | Description |
|---|---|
| AHSS | advanced high strength steels |
| BCC | body-centred cubic |
| BCT | body-centred tetragonal |
| BI | banding index |
| BIT | bainite isothermal transformation, also overaging or bainite holding (temperature) |
| BIW | body-in-white |
| CA | cellular automata (modelling) |
| CCD | charged-couple device |
| CP | complex phase (steel) |
| CR | cold-rolling/ (as) cold-rolled |
| DP | dual phase (steel) |
| EBSD | electron back-scattered diffraction |
| ELO | zinc coated (electroplated) |
| FCC | face-centred cubic |
| FE | finite-element (modelling) |
| GA | zinc coated (galvannealed) |
| GI | zinc coated (galvanised) |
| GND | geometrically necessary dislocation(s) |
| HDG | hot-dip galvanising |
| HEXRD | high-energy X-ray diffraction |
| HM | mean distance between bands |
| HR | hot-rolling/ (as) hot-rolled |
| IA | intercritical annealing (time/ temperature) |
| MP | multi-phase (steel) |
| pXRD | powder X-ray diffraction |
| TRIP | transformation induced plasticity (steel) |
| TADP | TRIP-assisted dual phase (steel) |
| XRD | X-ray diffraction |
| XRF | X-ray fluorescence |
| +Z | zinc coated (generic) |

## 1.    Introduction

In the drive towards higher strength alloys, a diverse range of alloying elements is employed to enhance their strength and ductility. Limited solid solubility of these elements in steel leads to segregation during casting which affects the entire down-stream processing and eventually the mechanical properties of the finished product. Although it is thought that the presence of continuous bands lead to premature failure, it has not been possible to verify this link. This poses as increasingly greater risk for higher alloyed, higher strength steels which are prone to centre-line segregation: it is thus vital to be able to predict the mechanical behaviour of multi-phase (MP) steels under loading.

This review lays the foundations of an experimental study to investigate the effects of centre-line segregation and banded microstructures on the mechanical behaviour of the phases present in a multi-phase steel comprising nominally 10% each of martensite, bainite and retained austenite, the remainder being ferrite. Due to the phase composition, this steel has been chosen as the ideal vehicle for understanding micro-mechanical behaviour of all of the major constituent phases of advanced high strength steels (AHSS). This research will make a significant contribution in enabling the generation of damage tolerant structures of MP steels for automotive applications and also the relevant material property data needed for next generation micro-mechanical models.

Cellular-Automata (CA) modelling was used to predict the rolling and cooling conditions necessary to produce samples with varying microstructures for the in-situ study. Micro-mechanical behaviour will be measured in the bulk and the centre-line of banded and non-banded specimens under tensile loading conditions using synchrotron radiation. Particular emphasis will be paid to strain partitioning between phases and progression of the strain induced transformation of the retained austenite to martensite.

This review covers the microstructure and properties of galvanised advanced high strength steels with particular emphasis to their use in automotive applications. In order to understand the origins of banding, the origins of segregation of alloying elements during casting and partitioning in the solid state will be discussed along with the effects on the mechanical behaviour and damage evolution under (tensile) loading. Attention will also be paid to the application of microstructural models in tailoring the production process to enable suppression of the effects of segregation upon banding. Finally, the theory and application of the experimental techniques used in this work to elucidate the structure and properties will be examined.

## 2.    On advanced high strength steels

In recent decades, governments in some of the world's major automotive markets, with a combined total of 70% of worldwide sales, have adopted varying forms of light-duty vehicle efficiency standards i.e. regulations requiring reduced $CO_2$ emissions, reduced fuel consumption rates, or increased fuel economy [1]. This trend looks set to continue, see Figure 1, from which it can be seen that in the first quarter of the 21st Century, the global target is to halve the allowable emissions from light vehicles such as passenger cars [2].

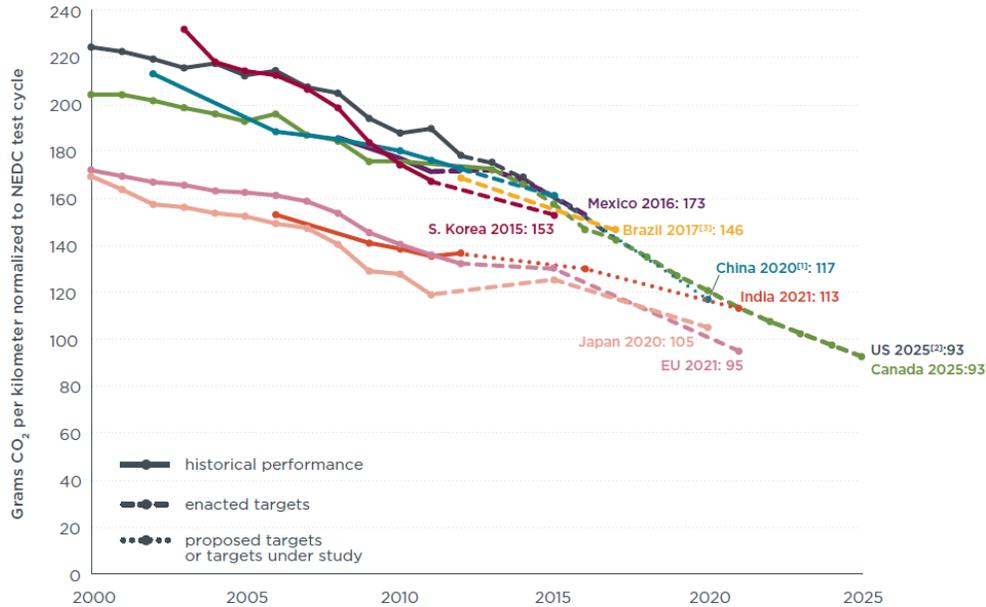

Figure 1: Comparison of global $CO_2$ regulations for passenger cars, in terms of gCO2/km. [1]

The regulation of emissions is seen as a key driver toward weight reduction as there is a direct correlation between mass and CO2 emissions [3]. For example, the European Parliament [4] has set a long-term target of 95 g/km $CO_2$ emission to be met by 2020 whereby the specific allowable emissions for each vehicle is based on the deviation from the average mass of new passenger cars in the previous three calendar years:

$$Specific\ emissions\ of\ CO_2(g/km) = 130 + a(M - M_0) \qquad (1)$$

Where:

$M$ = mass of the vehicle

$M_0$ = average mass of new passenger cars in the previous three calendar years

$a$ = 0.0457

On the other hand, ever increasing demands on automobile safety, in accident prevention and passenger protection, have led to an increase in average vehicle mass of 62 kg between 1970 and 2010 due solely to safety related equipment whilst the base body weight remained fairly constant for most of the same period [5]. The impact of these requirements is that materials should either lead to a reduction in vehicle mass for the same safety performance or an improvement in safety performance for the same mass. AHSS have thus been developed to meet both of these criteria and body weight savings of up to 35% are possible in combination with optimised forming and joining techniques [6-8].

The drive to reduce vehicle weight without compromising safety has led to the development of new grades of steel, partly since the controlling factor for the majority of automotive applications is the strength and/or Young's modulus, in this case measured by the tensile test, see Table 1. In the case of the automotive industry, the body in white (BIW[a]) is 60% gauged for stiffness and 40% for strength. In order to reduce weight then, either the modulus has to be increased or the tensile strength. Although there have been recent advances in high Young's modulus steels [9, 10], these are not at present commercially viable products. Therefore, for the purposes of this work the Young's modulus (of steel) will be assumed to be constant (around 200 GPa).

Table 1:    Required properties and main controlling factors of various vehicle parts. [7]

| Part | | Required properties | | | | |
|---|---|---|---|---|---|---|
| | | Panel rigidity | Dent resistance | Member rigidity | Fatigue strength | Crash strength |
| Outer panels | Door outer, etc. | ◉ | ○ | | | |
| Inner panels | Floor, etc. | ◉ | | ○ | ○ | ○ |
| Structural parts | Front rail, rear pillar, etc. | | | ○ | ○ | ◉ |
| | Front side member, side sill, etc. | | | ◉ | ○ | ◉ |
| | Door reinforcement, etc. | | | ○ | ○ | ◉ |
| Underbody parts | Suspension arm, disc wheel, etc. | | | ◉ | ◉ | |
| Main controlling factors apart from thickness of steels | | Young's modulus | Yield strength | Young's modulus | Tensile strength | Tensile strength |

Figure 2 shows the processing route for producing AHSS strip, which include hot-rolling, pickling, cold-rolling, annealing and coating. It should be noted that whilst AHSS can be produced in the as-hot-rolled condition [11-13], this work will concentrate on cold-rolled and galvanised strip steels. AHSS can be provided with or without a coating, usually zinc for the superior cathodic protection provided.

The choice of (zinc) coating applied is dependent upon the combination of coating line and the incoming substrate. In the case of hot-rolled steels a coating can be applied by dipping the steel into a bath of molten zinc i.e. hot-dip galvanising (HDG) or by electroplating. Due to the production costs, the latter process is usually only employed when it is not possible to apply the zinc via HDG due to poor wetting of the substrate.

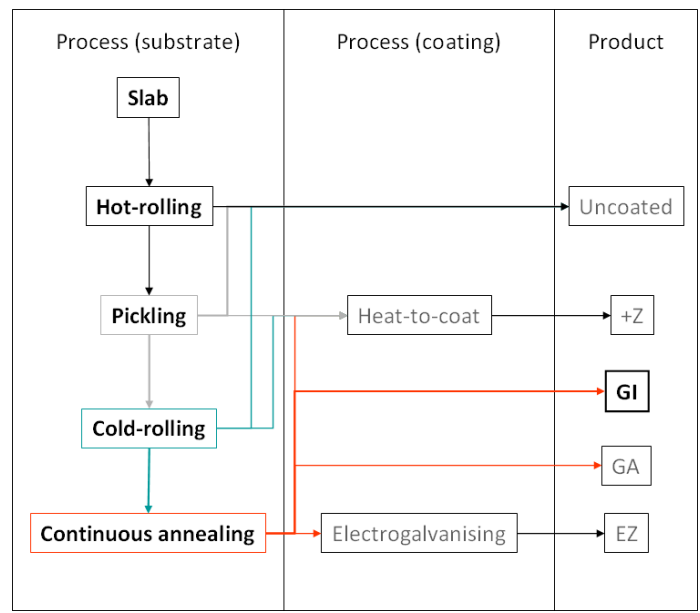

Figure 2:    Process routes for hot- and cold rolled AHSS with and without (zinc) coating. Courtesy of Tata Steel.

---

[a] BIW is the load-bearing structure of a vehicle excluding 'hang-ons' e.g. doors and bumpers etc.

For steels in the as-cold rolled condition, i.e. where a thinner gauge is necessary than can be achieved using hot-rolling alone, the steel strip is generally subjected to a thermal treatment in a continuous annealing line which may or may not include HDG. In the case of AHSS for automotive applications, this is generally the case and the thermal cycle is chosen to enable hot-dip galvanising during the cooling; either before martensite transformation, as in the case of DP, or after bainite transformation, as in the case of CP, TRIP and TADP. Figure 3 shows the influence of the major alloying elements on the phase transformations in AHSS.

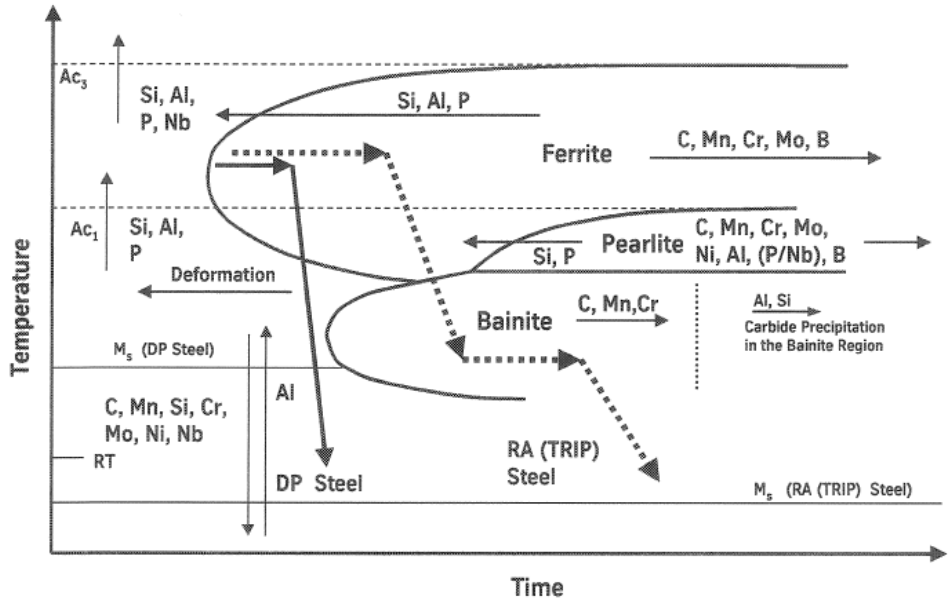

Figure 3:    Influence of alloying elements on TTT behaviour (schematic). After Ehrhardt [14]

Much of the traditional development (of AHSS) has concentrated on improving mechanical properties as measured by tensile tests, but there are not many applications in which the applied load path (either during forming or in service) is identical to that of the tensile specimen. In fact, high strength steel development is beginning to differentiate a number of classes of multi-phase (MP) steels based on real world applications [6, 15]. The first generation AHSS possess primarily ferrite-based microstructures [16] and are grouped according to the predominant microstructure type: dual phase (DP), complex phase (CP) and transformation induced plasticity (TRIP). More recently, hybrid TRIP-assisted DP (TADP) [14, 17] steels have been developed based on DP type microstructure i.e. ferrite-martensite, but with optimised chemistry and process for retention of a small percentage of retained austenite in the microstructure, see Figure 4.

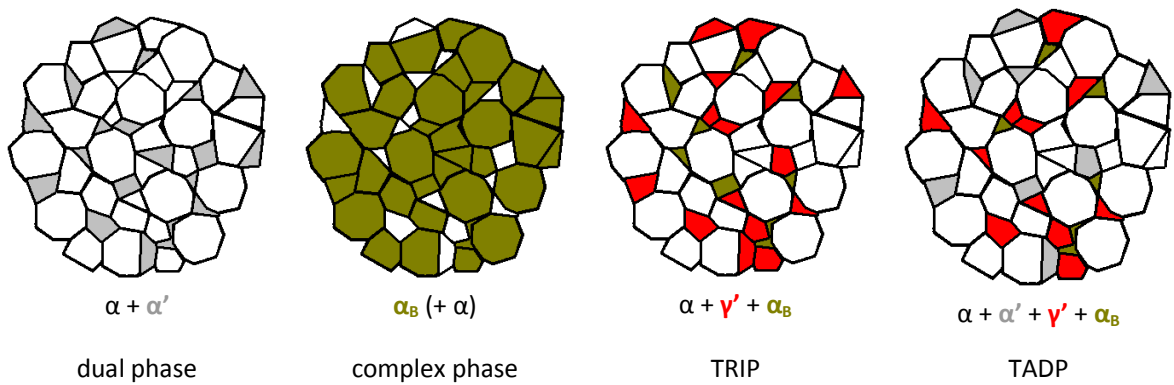

| α + α′ | αB (+ α) | α + γ′ + αB | α + α′ + γ′ + αB |
|--------|----------|-------------|------------------|
| dual phase | complex phase | TRIP | TADP |

Figure 4: Schematic representations of AHSS microstructures. After World Auto Steel [18]

## 2.1 Dual-phase and complex-phase steel

The strength of DP steels is largely dependent upon the relative hardness and volume fraction of the secondary phase (martensite) [19]; mechanical properties are characterised by low yield ratios, continuous yielding, high initial work hardening rates and a good combination of strength and ductility. DP steels are characterised by progressively increasing amounts of non-contiguous martensite in a ferrite matrix, whereby the strength increase of the steel is directly proportion to the amount of martensite present, see section 4.2.

During continuous annealing, Figure 5, the strip is annealed in the intercritical region (α+γ) i.e. above $A_{c1}$ and below $A_{c3}$ for a certain time, followed by quenching to below the martensite start temperature, whereby the high carbon austenite transforms to martensite islands (second phase) within a soft ferrite matrix. The cooling rate affects the amount of martensite produced. This is an important aspect as the volume fraction and strength of martensite within the structure are crucial for the high strength of the dual phase steel whereas the soft ferrite matrix leads to low yield strength [19]. The advantages over conventional high-strength low alloy steels are low yield stress to tensile strength ratios and a high formability [20].

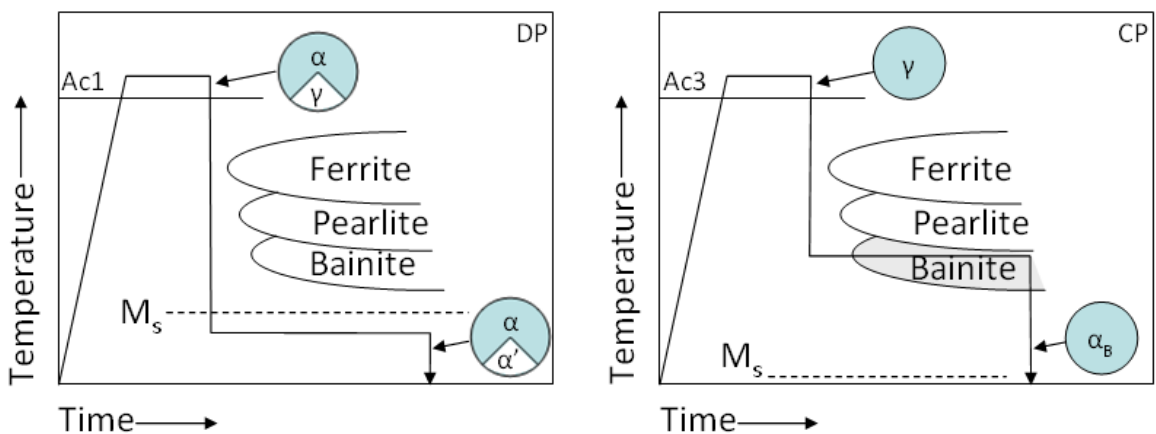

Figure 5: Schematic representation of typical annealing cycles of DP and CP steels. After Bleck [21]

Since the $M_s$ temperature for DP steels is generally below the zinc bath temperature (typically 460-470°C), the strip first has to undergo HDG prior to quenching. Cr is added to DP steels to avoid bainite formation during this treatment [21-23]. From Figure 3 it can be seen that Cr is a desirable element as it stabilises austenite during cooling; the general effect in DP steel is to increase the bainite region thereby reducing the kinetics of transformation.

CP steels on the other hand are annealed in the fully austenitic phase region to allow homogenisation of carbon to the equilibrium composition. The carbon content of CP steels is chosen to suppress ferrite transformation during quenching to a temperature within the bainite region. Isothermal holding at the bainite holding temperature, followed by rapid quenching to room temperature results in a fully bainitic microstructure.

## 2.2    TRIP and TRIP-assisted steel

The group of TRIP steels is named after the inherent dominant TRansformation Induced Plasticity effect in those steels. Fischer identified two types of TRIP steel, namely [24]:
- High-alloy (stainless steel) containing substantial amounts of Ni and Cr leading to a meta-stable austenitic steel at room temperature, and
- Low-alloy steel consisting of a ferritic-bainitic matrix with 10-15% retained austenite in the form of small packets in the parent phase grains or as layers along the grain boundaries.

The latter type of steels are most commonly used for automotive applications and are produced in a similar annealing cycle to DP steels, see Figure 6. The process route is designed to enrich austenite with carbon to ensure room temperature meta-stability. TRIP steels consist of a ferrite-bainite microstructure with retained austenite. Under loading metastable austenite undergoes diffusionless transformation to martensite, the so-called TRIP effect, and thereby extending uniform plastic elongation; see also sections 3.3 and 4.3. At the same time the martensitic transformation leads to increased strength via composite strengthening.

To retain the austenite in the final microstructure, an isothermal bainitic hold stage in the overaging section during annealing plays a key role. Aluminium and/or silicon are added to retard carbide precipitation during the bainite formation. Well balanced alloying and processing is essential for proper stabilisation of austenite during the bainitic hold in order to prevent a premature transformation into martensite during the final cooling to room temperature. All these aspects lead to the unique multiphase structure of ferrite, bainite and dispersed metastable retained austenite in TRIP steels [19].

Typical TRIP steel annealing processes after cold rolling involve a soak in the temperature range 300 to 450°C to allow for bainitic transformation. Retained austenite is obtained mainly by chemical stabilisation through carbon enrichment which is achieved during cooling from the intercritical temperature through ferrite re-transformation (to a limited extent) and mainly through bainite formation. Fast bainitic transformation kinetics is required to match the relatively short time available during hot dip galvanising. In the case of hot dip galvanizing lines without an overaging section, a relatively slow cool from the intercritical temperature can be used to promote carbon enrichment of the austenite due to ferrite and bainite formation.

In both routes, cementite precipitation must be prevented to ensure the optimisation of carbon enrichment in the austenite, which requires significant alloying with carbide inhibitors such as Si and Al [25].

Due to the insolubility of Si in cementite, diffusion of Si atoms away from the growing cementite nuclei becomes the rate determining process leading to retardation of cementite precipitation kinetics [26].

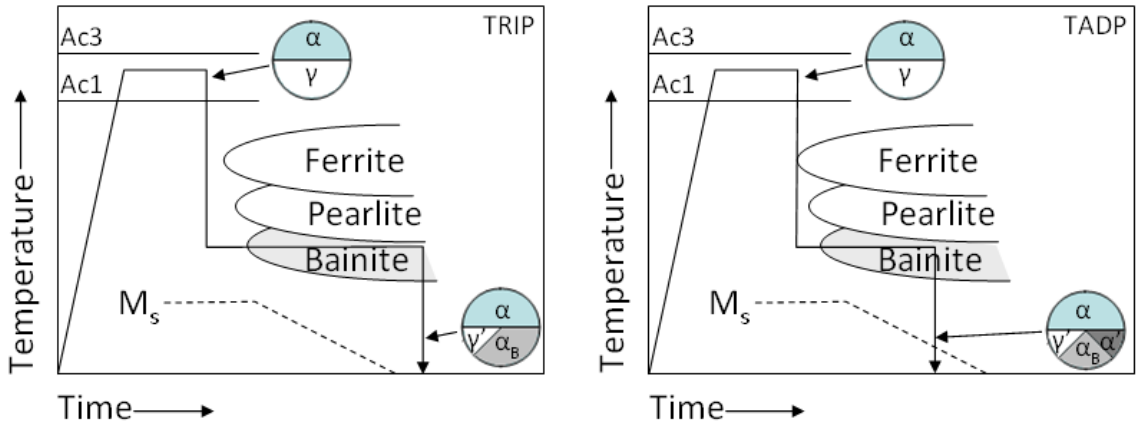

Figure 6: Schematic representation of typical annealing cycles of TRIP and TADP steels. Adapted from Bleck [21]

Commercial cold rolled annealed TRIP steels are typically based on a 0.2C-1.5Mn-1.5Si composition, see Figure 7 [27]. The maximum tensile strength of such a material is 800 MPa with a total elongation up to 30%. Although Si leads to higher strength due to solid solution strengthening, it also results in poor surface quality. Moreover, high Si contents cause difficulties during hot rolling and hot-dip galvanising due to the formation of a fragile intermetallic layer related to the presence of relatively stable $Mn_2SiO_4$ oxides on the strip surface [28, 29].

The higher Si concept has been replaced over the years with a partial or complete substitution with Al, with the CMnAl chemistry alleviating the issue of bad Zn wetting to some extent depending on the alloy content. Addition of Al leads to a much faster bainitic reaction and prevents carbide precipitation but it has a weaker solid solution hardening effect and it readily forms $Al_2O_3$ oxides causing difficulties in casting and possibly during hot dip galvanizing as well.

Most of the literature on low Si TRIP steels focuses on partial replacement of Si with Al that is also known as a cementite inhibitor but is not expected to influence coatability [30]. The great advantage of Al over Si in the case of hot-dip galvanising is that it is the only element apart from Co known to accelerate bainite kinetics. Unlike Si, Al is not a solid solution strengthening element: considering full replacement of Si with Al in the above mentioned 0.2C-1.5Mn-1.5Si base composition, a loss in tensile strength of about 100MPa can be expected [31]. Moreover although it seems to be the most promising element to replace Si, Al is also a strong nitride former that would affect expected precipitation strengthening from V, Nb or Ti additions.

The only other well-known cementite inhibitors are copper [32, 33] and phosphorus [33-35]. Cu is an austenite stabilizing element offering similar solid solution strengthening to Mn [31]. It is used in combination with Ni in order to avoid the hot-shortness phenomenon and can partially replace Si [32, 36].

P additions provide significant solid solution strengthening, i.e. about 680 MPa/wt.% [31] but in order to avoid segregation, the P content should be kept below 0.1 wt.% [35-37]. This is not sufficient to totally replace Si or Al with respect to cementite precipitation but allows Si to be decreased below 0.6 wt.% which is regarded as the critical upper limit of Si content, above which it interferes with zinc wetting during galvanising[36]. Although P adds strength and ductility (through increase in γ_r fraction), its application is limited in automobile applications due to poor weldability [38].

Figure 5 shows roles of these alloying elements to achieve the right combination of strength and good galvanizing surface properties. Though both Si and Al have high free enthalpy of oxide formation, it is the location of the oxidation which controls the zinc coating characteristics. The enrichment response of the two elements under normal annealing conditions varies with Si tending towards external oxidation and covering the as-annealed surface while Al in comparison though oxidizing but internally [27].

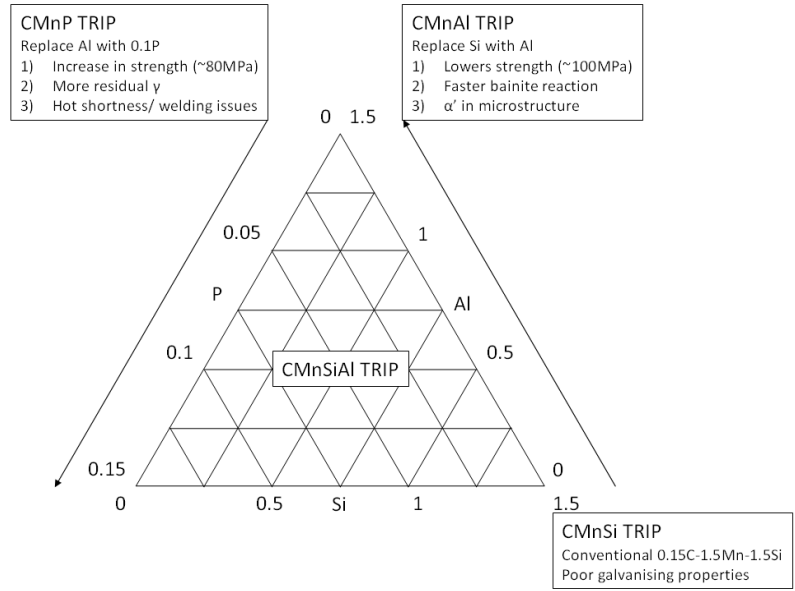

Figure 7:   Role of chemistry concepts for strength vs. coating characteristics of TRIP steels. After Bode [27]

In order to increase the strength to 1000 MPa without increasing the carbon content (in order to maintain the weldability), microalloying techniques are employed to introduce precipitation strengthening [39] e.g. Ti and/or Nb additions and V additions with high nitrogen content. The effect of Nb is complex, because while it delays bainite transformation, either by enhancing ferrite formation as a result of grain refinement [40] or by blocking nucleation sites, it does suppress carbide formation during overaging [21, 40-42].

Although TRIP steels have been applied successfully, their application is still very limited due to issues with weldability, crash performance and coatability. Furthermore, whilst the DP and TRIP steels have excellent plane strain formability and stretchability due to extended work hardening capacity, they do have a number of critical disadvantages: edge cracking sensitivity and, as is becoming increasingly apparent, they exhibit low damage tolerance due to problems of load transfer between hard and soft phases; the greater the difference in strength between the phases, the less likely it will be that they will co-deform [20].

A hybrid TRIP-assisted DP steel has been recently developed based on DP type microstructure i.e. ferrite-martensite, but with optimised chemistry and process for retention of a small percentage of retained austenite in the microstructure [17]. Al is coupled with a lower amount of Si to provide solid solution strengthening whilst not reducing the coating capability. This results in multi-phase steel comprising of ferrite and around 10% each of martensite, bainite and metastable austenite. This steel forms the basis for this research work.

## 3. Origins of banding in AHSS

In the drive towards higher strength alloys, a diverse range of alloying elements is employed to enhance either the strength or the ductility or both. Limited solid solubility of these elements in steel leads to segregation during casting which affects the entire down-stream processing and eventually the mechanical properties of the finished product via the introduction of banding in the microstructure. Although it is thought that the presence of continuous bands lead to premature failure, it has not been possible to verify this link. This poses as increasingly greater risk for higher alloyed, higher strength steels which are prone to centre-line segregation: it is thus vital to be able to predict the mechanical behaviour of multi-phase (MP) steels under loading.

### 3.1 Segregation during casting

During solidification, solute is partitioned between the solid and liquid to either enrich or deplete interdendritic regions. This naturally leads to variations in composition on the scale of micrometres, i.e., to micro-segregation. Macro-segregation, however, refers to chemical variations over length scales approaching the dimensions of the casting, which for large ingots may be of the order of centimetres or metres. Micro-segregation can be removed by homogenisation heat treatments, but it is practically impossible to remove macro-segregation due to the distances over which species are required to move.

Solute atom redistribution during dendritic solidification is driven by the equilibrium partitioning of chemical elements within the liquid-solid phase field. The two-phase field is defined as a function of temperature by liquidus and solidus curves of the equilibrium phase diagram.

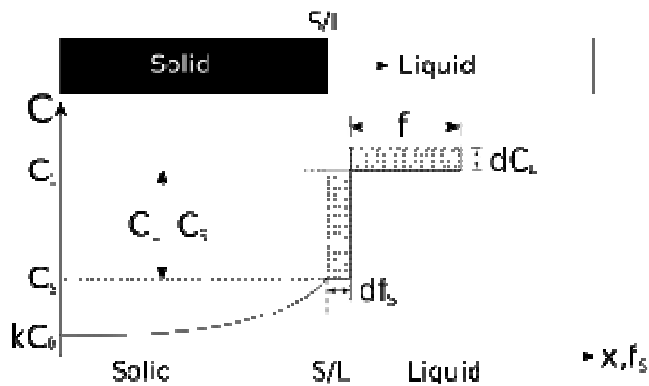

Figure 8: Solidification scheme used for deriving the Scheil-Gulliver equation [43].

At any given temperature, for an alloy with bulk solute concentration $C_0$, the solute concentration of the solid may be designated as $C_S$ and that of the liquid as $C_L$, as defined by a tie-line at that temperature, see Figure 8. The hatched areas in the figure represent the amount of solute in the solid and liquid. The redistribution or partitioning of solute is then defined as the equilibrium partition ratio, k, as [44]:

$$k = \frac{C_s}{C_L} \tag{2}$$

Mass conservation dictates that the solute redistribution in the solid must come from the liquid, therefore the weight fractions in the solid($f_s$) and the liquid ($f_L$) must add up to unity, thus:

$$f_s + f_L = 1 \tag{3}$$

In which case the mass balance may be rewritten as:

$$C_L(1-k)df_s = (1-f_s)dC_L \tag{4}$$

Integrating using the boundary condition,

$$C_L = C_0 \quad at \quad f_s = 0 \tag{5}$$

yields the Scheil-Gulliver equation [45] for the composition of the liquid during solidification, thus:

$$C_L = C_0(f_L)^{k-1} \tag{6}$$

And for the solid:

$$C_s = kC_0(1-f_s)^{k-1} \tag{7}$$

The Scheil-Gulliver equation is based on a number of simplifying assumptions,[46] including negligible under-cooling, complete diffusion in the liquid in the volume element, negligible diffusion in the solid and constant $k$ throughout solidification. Other more complicated equations of solute distribution, including the effects of diffusion in the solid and convection, have been evaluated [44] but Eq. (4) adequately demonstrates solute enrichment in the solid as solidification proceeds. The first solid to solidify has the lowest solute content and the concentration of solute, or solutes in multi-component alloys such as steels, increases in the solid phase with increasing solidification, with the highest concentration of solutes in the last portion of liquid to freeze.

Values of $k$, assumed to be independent of temperature, of some elements commonly found in steels have been reported in the literature using direct measurement of concentration profiles, see Table 2 [47] or by using thermodynamic modelling software such as THERMO-CALC,[b] see Table 3 [48] These values seem to be in good agreement with the experimental values, especially in the more abundant elements such as Mn and Ni, although on the basis of the partition coefficients in Table 2, it would appear that aluminium would not segregate significantly during solidification; this is almost certainly due to the residual levels of aluminium (<0.04 wt. %) reported in the study .

---

[b] THERMO-CALC is a trademark of Thermo-Calc, Stockholm

Table 2:  Equilibrium partitioning coefficients for common elements in steel. After Fisher [47]

| Element | k |
|---|---|
| P | 0.14 |
| Nb | 0.23 |
| Cr | 0.33 |
| Mn | 0.71 |
| Ni | 0.83 |

Table 3:  Maximum partition ratios of three different HSLA steels. After Chakrabarti [48]

| Slab No. | Slab composition (wt.%) | | | | | | | | | |
|---|---|---|---|---|---|---|---|---|---|---|
| | C | Si | Mn | P | S | Ni | Al | Nb | Ti | V |
| 1 | 0.10 | 0.31 | 1.42 | 0.017 | 0.005 | 0.32 | 0.045 | 0.045 | 0.002 | 0.052 |
| 2 | 0.10 | 0.28 | 1.41 | 0.013 | 0.001 | 0.30 | 0.029 | 0.027 | 0.001 | 0.05 |
| 3 | 0.09 | 0.38 | 1.52 | 0.011 | 0.002 | 0.51 | 0.036 | 0.036 | 0.007 | - |
| Ave. k | 0.26 | 0.63 | 0.71 | 0.21 | 0.02 | 0.84 | 1.02 | 0.15 | 0.19 | 0.58 |
| Δk | <0.01 | 0.01 | <0.01 | 0.03 | <0.01 | 0.07 | n/a | 0.01 | 0.02 | 0.03 |

Solute elements with low values of $k$ have the greatest tendency of segregate i.e. phosphorus has a very strong tendency to segregate during solidification. However, the amount of the element present is also a factor. Therefore, Mn, generally present in much higher concentrations than P, is a more significant factor in segregation and banding than P despite its higher $k$ value. E.g. according to Eq. (7) , for a 1.0 wt.% Mn steel, Mn would vary from 0.70 wt.%  at the beginning of solidification to 1.60 wt.% at the end of solidification. The effect of this partitioning behaviour was demonstrated by Black [49] who measured segregation across the diameter of a high carbon steel bar, see Figure 9. The variation in Mn content varies from between 0.5 wt.% to 1.4 wt.% Mn, which is accordance with the theory as outlined above.

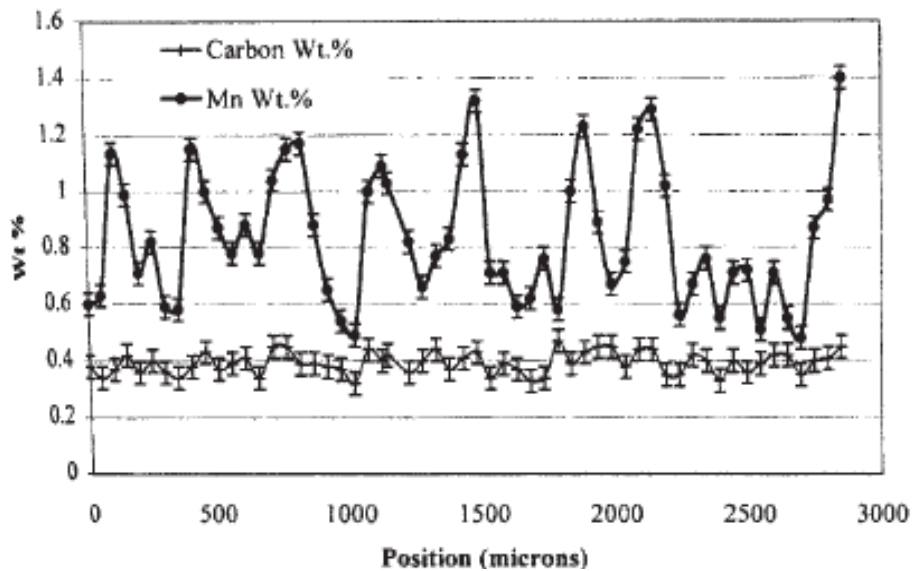

Figure 9:   Variations in Mn and C concentrations across a 95 mm diameter, quenched and tempered, 4140 steel bar containing 1 wt.% carbon [49].

Pickering has recently performed an extensive survey of the available literature on micro- and macro-segregation and the attempts to model and ultimately control this phenomenon [50]. Although his work focuses on steel ingots which have a significantly different cooling trajectory to continuously cast steel slabs, the basic principles can be applied. In particular the formation of V-segregation and the resultant centre-line segregation as a result of enrichment of the final material to solidify can be transposed to the solidification of slabs. In the case of slabs, the shear planes would not be caused by the metallostatic head, but by the shear imposed across the solidifying region due to friction due to movement along the mould wall. One way of combating the centre-line segregation in continuous casting is the introduction of pinch rolls in the strand ahead of complete solidification to squeeze the still liquid region back up the strand where the solute can be redistributed in the liquid head. This requires accurate alignment of the casting machine to avoid entrapment of liquid pockets which will remain high in solute concentration. Furthermore, this has little or no effect on micro-segregation and leads to a gradual enrichment along the strand [51].

Pickering also gives details of the difficulties faced by modellers in capturing the segregation due to complexity of each domain (liquid, solid and mushy) and the interaction between them. For example the Scheil equation outlined previously is an expression of the redistribution of solute atoms in equilibrium and does not take into account solute redistribution as a result of fluid flow and shrinkage. Pickering suggests a method of detecting segregation using x-ray fluorescence [52], although this is limited in the case of carbon which cannot be detected with this technique. He proposes to accurately measure carbon in a few areas and infer the content in the remaining structure based on the segregation profile. This measurement is in turn heavily dependent upon an accurate measurement of carbon, in itself no mean feat.

What is clear is that if reduction of macro-segregation is difficult, reduction of micro-segregation is more difficult still as this requires controlling dendritic formation and growth [53, 54]. The are two ways in which to reduce the degree of micro-segregation during solidification: steady-state planar growth with equilibrium partitioning at low growth rate with a positive temperature gradient or solidification at growth rates exceeding the limit of stability such as is achieved by fast cooling to a temperature below the liquidus i.e. a high degree of undercooling [55]. In the former case there is no reduction of centre-line segregation as this approaches the Scheil condition i.e. liquid enrichment as a function of the partitioning coefficient.

Kurz *et al.* [53, 54] propose a model for the under-cooling at solid-liquid interface $\Delta T$ as the sum of the under-cooling due to: solute diffusion ($\Delta T_C$), Gibbs-Thomson curvature ($\Delta T_R$), kinetics ($\Delta T_k$) and the interaction between cells and the solute distribution ahead of cell tip under the uni-directional solidification condition ($\Delta T_U$).

$$\Delta T = \Delta T_C + \Delta T_R + \Delta T_K + \Delta T_U \qquad (8)$$

Where:

$$\Delta T_C = \frac{k\Delta T_0 Iv(P)}{1-(1-k)Iv(P)} = \sum_i m_i (C_{L,i}^{tip} - C_{0,i}) \qquad (9)$$

$$\Delta T_R = \frac{2\Gamma}{r} \qquad (10)$$

$$\Delta T_K = \frac{V}{\mu_0} \qquad (11)$$

$$\Delta T_U = \frac{GD}{V} \qquad (12)$$

Where $k$ is the equilibrium partitioning coefficient, $Iv(P)$ is the Ivantsov function, $m_i$ is the liquidus slope of element i, $C_{L,i}^{tip}$ the concentration of element i in liquid at dendrite tip, $C_{0,(i)}$ the initial solute concentration (of element i), $\Gamma$ the Gibbs-Thomson coefficient ($=\sigma /\Delta S$), $r$ the dendrite tip radius, $V$ the growth rate, $\mu_0$ the linear kinematic coefficient ($=V_0 V_m \Delta S/RT_f$) and $G$ and $D$ the temperature gradient and diffusion coefficient in liquid respectively.

Although undercooling is effective in reducing the primary arm spacing [53, 55, 56], segregation itself cannot be completely eliminated, only the scale at which it takes place. Therefore, for practical applications, one has to assume the presence of segregation in the as-cast microstructure and attempt to deal with the effect on the local transformations in down-stream applications. As if partitioning in the liquid were not enough to contend with, increasingly advanced modelling and experimental techniques have illustrated that partitioning not only influences transformation, but may also be the result of it, albeit it a different length scale, as will be discussed in the following section.

## 3.2 Partitioning/ diffusion induced grain boundary migration

Since we have established in section 2 that the microstructures of multi-phase steels are the result of quenching in the intercritical region, it follows that differences in local chemistry will have an effect on the transformation products and ultimately the local and bulk mechanical properties. Wycliffe showed that there is a concentration gradient across the martensite grain away from the grain boundary with ferrite, see Figure 10. This suggests that there is partitioning in the solid state, even at the very short times experienced in industrial annealing processes and increases as function of time [57].

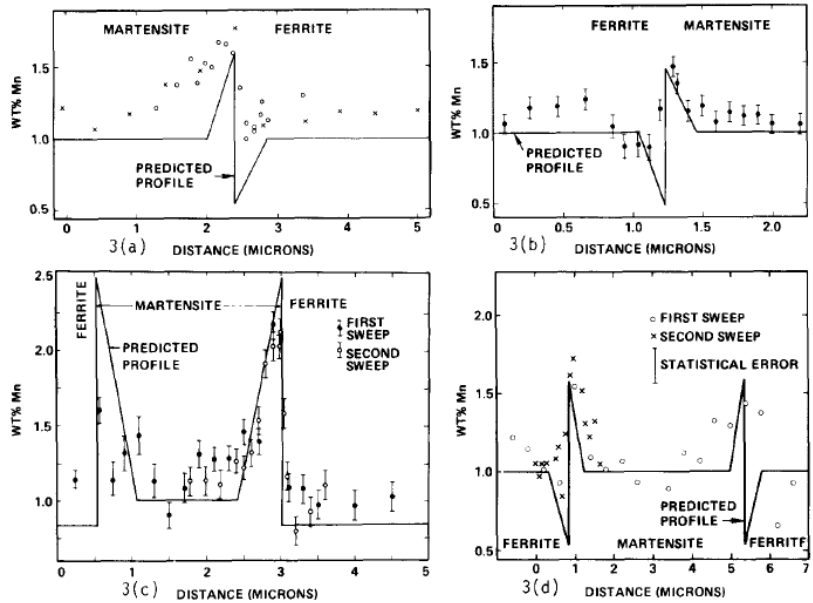

Figure 10: Predicted and observed Mn content across ferrite-martensite grain boundaries in dual phase steels heat-treated at 750°C for: (a) 500 s, (b) 100 s, (c) 50 h and (d) 500 s [57].

This conclusion is supported by Navara and co-workers who concluded that the initial stage of austenite formation in dual-phase steels is related to manganese diffusion induced ferrite boundary migration [58]. This process depends on the presence of oriented grain boundary dislocations which enhance manganese diffusion and provides the right conditions for austenite nucleation. Subsequent austenite growth is thus dependent initially on continued boundary migration, but at a later stage on manganese diffusion through ferrite, as well as some redistribution within the growing austenite. Toji et al however found no evidence of grain boundary migration of either Mn or Si during intercritical annealing and concluded that austenite formation was dependent upon carbon, although Mn was instrumental in retarding the retransformation from austenite to ferrite [59].

Barbé demonstrated that in the early stages of intercritical annealing in industrial galvanising lines, that the transformation to austenite is indeed controlled by carbon diffusion, as the diffusion of substitutional elements is too slow. This leads to a build-up of ferrite stabilising elements (e.g. Si) on the ferrite side of the transformation front, and austenite stabilising elements on the other side, see Figure 11 [60]. The role of Al was not investigated, but as a ferrite stabiliser it would also be expected to partition into the ferrite in the same way as Si.

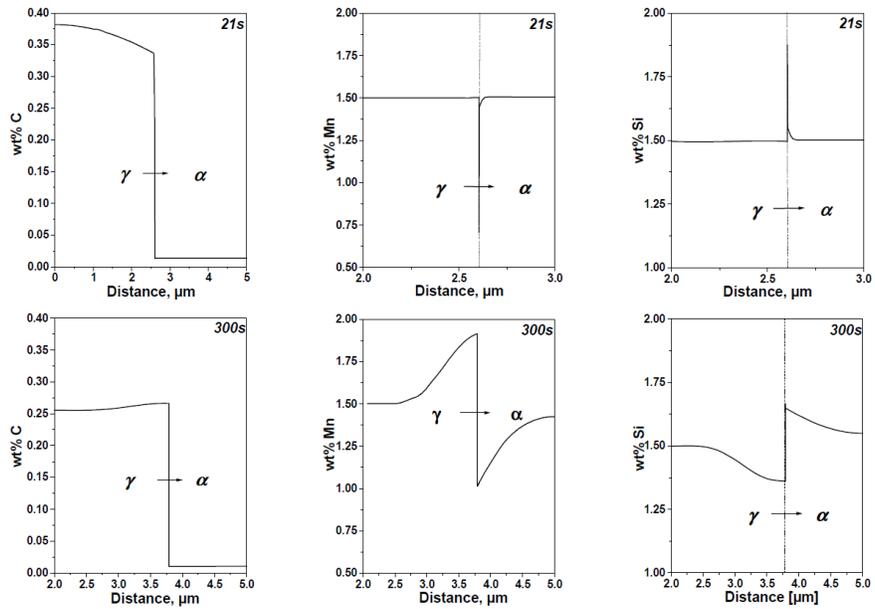

Figure 11:    Calculation of C, Mn and Si profiles across the austenite to ferrite transformation front (shown by arrow) at different intercritical annealing times [60].

Kumar [61] also found evidence of Mn partitioning during heat-treatment which enhanced hardenability and thus formed martensite. The volume fraction of martensite and the manganese redistribution were found to be dependent on both intercritical annealing temperature and holding time. Dispersed martensite was obtained at higher annealing temperatures in the second stage, while ring or island-type of martensite was seen at relatively lower annealing temperatures. Although the findings are not disputed, Mn diffusion was only detected at the longest annealing time 2-15 h, which is far outside of industrial processing practice.

More recently Dmitrieva and co-workers have measured concentration profiles across martensite-austenite phase boundaries using Atom-Probe Tomography [62]. Figure 12 quite clearly shows austenite and phase boundary enrichment of Mn. The phase boundary is also depleted in Al and Si which concurs with the relative partitioning coefficients and the observations from Barbé [60].

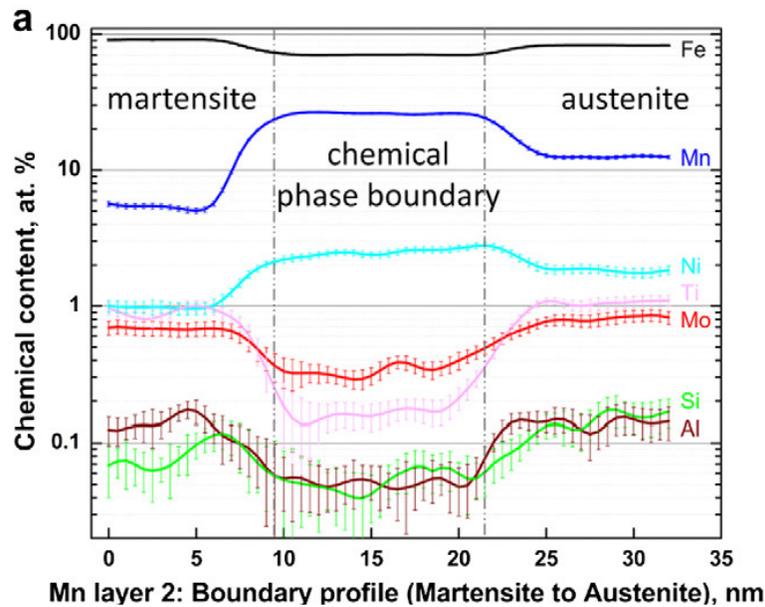

**Mn layer 2: Boundary profile (Martensite to Austenite), nm**

Figure 12: Concentration profiles across martensite/austenite phase boundary [62].

In the same paper they propose a model in which Mn mobility is linked to diffusivity differences between martensite and austenite lead to a Mn flux towards the retained austenite. Since the diffusivity in austenite is lower, it cannot accommodation the flux, leading to a build-up of Mn at the phase boundary, the composition of which is given by local equilibrium.

### 3.3 The martensite transformation

Interstitial iron-based martensites are supersaturated solid solutions of interstitial atoms, most commonly carbon and nitrogen in a body centre cubic iron lattice which is dominantly strained along the octahedral interstices along the c direction to form a body-centre tetragonal lattice [63, 64]. The martensite transformation can occur under two conditions [21]:

(1)        The applied temperature is below the $M_s$-temperature: The austenite phase is thermodynamically unstable and transforms to martensite due to cooling without deformation.

(2)        Stress is applied: The deformation facilitates a martensite transformation of the retained austenite above the $M_s$-temperature after reaching a certain threshold value.

The martensitic transformation of retained austenite due to an applied stress is the underlying mechanism of the TRIP effect. During the diffusionless transformation the crystal structure changes and is accompanied by a volume and shape change. This change in shape and volume during transformation increases the plastic strain and the transformation of the softer austenite to harder martensite provides superior work-hardening behaviour [65] This can help to delay the start of local necking as well as stopping the extension of cracks. However, the hard martensite particles created during the transformation can have a negative effect as they can serve as void nucleation sites [66].

### 3.3.1    Kinetics of and the effect of alloying additions on thermally activated martensite formation

In MP steels austenite transforms to martensite during cooling due to low austenite stability. Transformation is dependent on the intercritical annealing temperature, the heating rate to the intercritical temperature as well as the carbon content in the austenite at the end of the intercritical annealing. In the absence of auto-tempering during quenching the carbon content and its spatial distribution in the intercritical austenite regions will be passed on to the final martensite islands. Additionally, at high heating rates the ferrite cannot fully recrystallise before reaching the intercritical temperature. Therefore, austenite mainly accumulates at elongated prior carbon rich regions e.g. pearlite colonies, followed by rapid growth into these areas. In contrast, a low heating rate gives the ferrite enough time to recrystallise completely and austenite nucleates predominantly at the recrystallised ferrite grain boundaries [20].

The volume fraction of martensite as a function of undercooling below the $M_s$ temperature can be described by the Koistinen and Marburger equation [67, 68], whereby the volume fraction of martensite, $f$, is assumed to be a function of undercooling below $M_s$ only:

$$f = 1 - e^{-\alpha_m (M_S - T)} \tag{13}$$

Where $T$ is the temperature and the rate parameter $\alpha_m$ is a function of composition, thus:

$$\alpha_m = 10^{-3} \left( 27.2 - \sum_i S_i x_i - 19.8(1 - e^{-1.56 x_C}) \right) \tag{14}$$

And

$$\sum_i S_i x_i = 0.14 x_{Mn} + 0.21 x_{Si} + 0.11 x_{Cr} + 0.08 x_{Ni} + 0.05 x_{Mo} \tag{15}$$

(where $x_i$ is the concentration in wt.%). If other phases such as ferrite are formed in an earlier stage of the austenite decomposition, $f$ in Eq. (13) is the normalised martensite fraction i.e. $f = 1$ when the maximum possible martensite fraction has been reached.

Experimental data [69] shows that the transformation rate just below $M_s$ is initially more gradual, which is related to the austenite grain size. The overall best fit to the experimental fraction curve is therefore obtained using a so-called theoretical martensite start temperature $T_{KM}$ that is somewhat lower than $M_s$, and is given by a modified Andrews' equation [68, 70]:

$$M_S(°C) = 565 - \sum_i K_i x_i - 600(1 - e^{-0.96 x_C}) \tag{16}$$

Where:

$$\sum_i K_i x_i(°C) = 31 x_{Mn} + 13 x_{Si} + 10 x_{Cr} + 18 x_{Ni} + 12 x_{Mo} \tag{17}$$

which describes how substitutional elements influence $M_s$.

In principle, martensite transformation is assumed to be instantaneous, which is why there is no explicit time dependence in Eq. (13). However, it can be argued that in reality martensite does need a finite amount of time to nucleate and grow. On the run-out-table of a hot strip mill, for example, the surface temperature can be considerably lower than the average strip temperature for a short time. It is not impossible that the time at the lower temperature is too short for martensite to form, but this is not known as an exact calculation of the heat of transformation of martensite is difficult. What is known, however, is that the stored energy in martensite is high, around 400 J/mol (7151 J/kg) [69]. This stored energy is mainly strain energy due to the mechanism of the transformation. The stored energy is subtracted from the usual transformation heat calculation of ferrite and cementite.

### 3.3.2    Ms and Ms<sup>σ</sup> temperature and the role in mechanically induced martensite

As the martensite start temperature ($M_s$) is used to describe the thermodynamic stability of austenite against cooling, recent work states that the $M_s^\sigma$ temperature characterises the point at which the phase transformation changes from stress-assisted to strain-induced [71]. At this temperature the yield strength of the parent phase equals the stress required to initiate martensite transformation featuring the highest uniform elongation. A lower $M_s^\sigma$ temperature refers to a higher stability of retained austenite due to the carbon enrichment at the bainite hold section and a higher $M_s^\sigma$ temperature leads to a drop of the stability of retained austenite. To meet the requirements of being in the strain induced region of the TRIP effect the $M_s^\sigma$ temperature should be below room temperature, otherwise a transformation of the austenite can occur during cooling or elastic deformation [72]. In Figure 13 the temperature ranges for the different interactions are schematically represented. Above the $M_d$ temperature no martensite transformation can be induced by deformation.

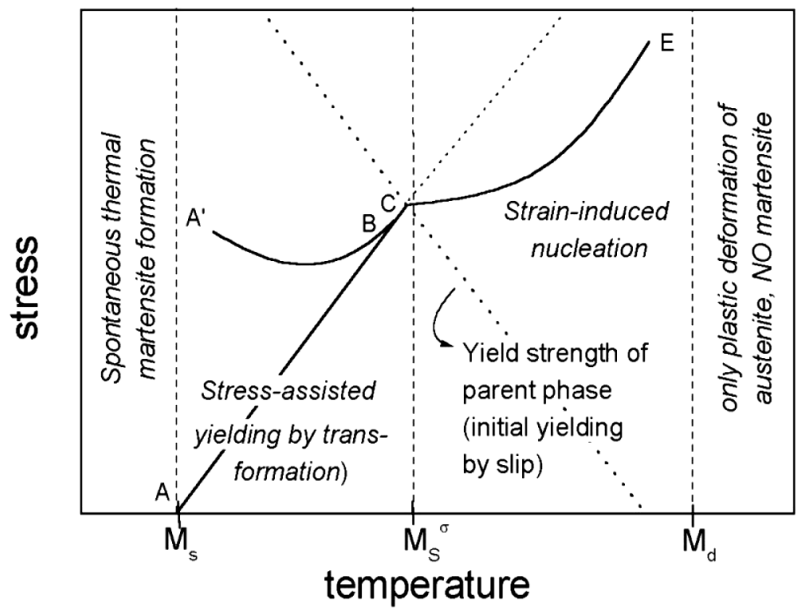

Figure 13:    Schematic representation of stress-assisted and strain-induced regimes of mechanically induced martensite transformation and definition of the $M_s^\sigma$ temperature [71].

### 3.3.3    Formation of retained austenite during continuous annealing

In TRIP-aided steels the stability of retained austenite is the most important parameter controlling the mechanical properties as mechanically activated transformation of this phase leads to martensite [71]. The stability of retained austenite is primarily dependent on the carbon content: at low carbon levels (<0.6 wt.% C), transformation to martensite occurs at low levels of strain and above 1.8 wt.% C it is assumed that no mechanically induced transformation occurs [66]. Additional factors are the size and shape of the retained austenite islands, concentration of substitutional alloying elements, crystallographic orientation as well as the strength of the retained austenite and surrounding ferrite and bainite phases [73, 74].

The volume fraction of retained austenite in TRIP steels varies as a direct result of bainite formation during bainite isothermal holding (BIT). As bainite forms, it expels carbon across the transformation boundary, thereby increasing the carbon concentration of the untransformed grain and, hence, increasing the stability of the austenite. Tuning the temperature and time of BIT allows optimisation of the volume fraction of austenite to be retained after quenching to room temperature, although at longer holding times bainite will replace the intercritical austenite and carbides can be formed. This would lead to a lower concentration of carbon in the austenite phase and consequently decrease the volume fraction of retained austenite in the final microstructure [66].

Figure 12 shows the effect of Si and Al on the volume fraction of austenite as a function of bainite holding time. It is apparent from the graph that Al leads in general to faster transformation kinetics and thus favours short holding times. Si on the other hand leads to slower transformation kinetics, but an overall increase in retained austenite fraction. Furthermore, the retained austenite fraction of the Si steel does not reduce with the BIT which Jacques attributed to the formation of carbides in the Al containing steels [73].

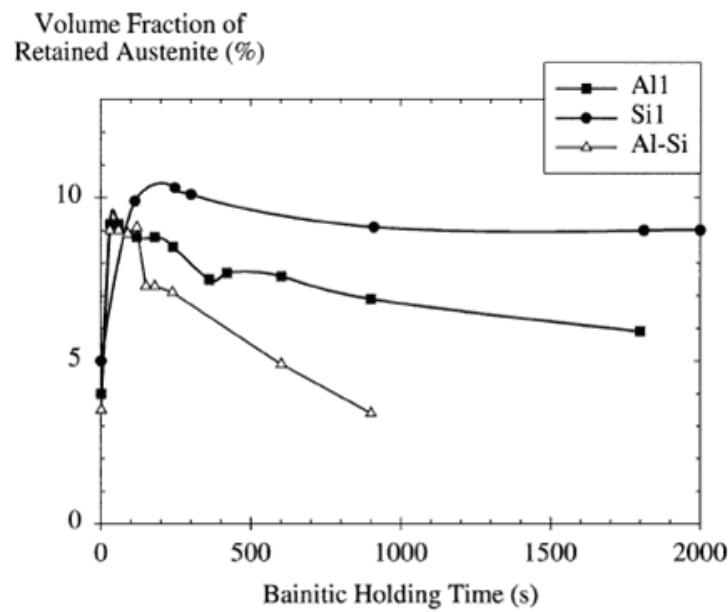

Figure 14:    Effect of bainitic holding time on volume fraction of austenite in Al and Si TRIP steels [73].

In cold-rolled material, the austenite grain size is much reduced due to a finer initial grain size. These grains are more stable because of a lower martensite start temperature due to higher carbon enrichment and a decreased number of martensite nucleation sites in each grain.

## 4.    The effect of segregation and banding on mechanical properties

Firstly, what do we mean by banding? The most basic description is the orientation of one or more microstructural feature in any given direction which for strip steels is usually, though not exclusively, parallel to the rolling direction.

For the purposes of this work, we will be concentrating primarily on banding caused by macro- and micro-segregation. Casual observation of the iron-carbon binary equilibrium diagram, see Figure 15, reveals that the carbon content has the most significant effect on segregation (in AHSS) due to the increase the solidification range with increasing carbon content. Alloying elements such as manganese and chromium remain dissolved in the liquid state, whilst silicon and aluminium partition to the solid phase [75]. Due to fluid flow phenomena in the liquid steel and inter-dendritic space the enriched liquid is washed from the inter-dendritic zone into the bulk liquid [76]. By those phenomena the bulk liquid is incrementally enriched with alloying elements. At complete solidification this leads to an enriched centre-line of the slab and thus centre-line banding.

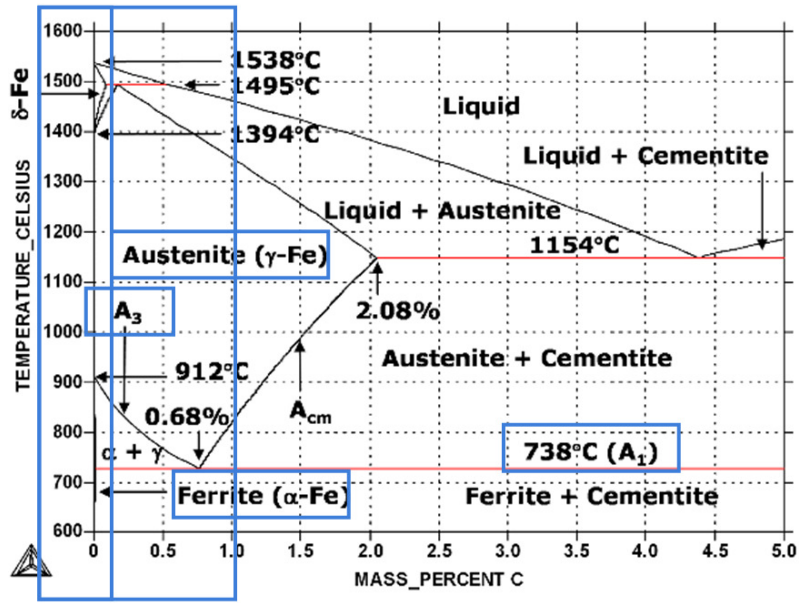

Figure 15:    Iron-carbon binary equilibrium diagram as calculated using THERMO-CALC

Figure 16 shows the degrees of centre-line banding in a typical multi-phase steel microstructure and although the banding can be detected visually, characterisation of banded microstructures has proved to be more complex than one would presume at first glance.

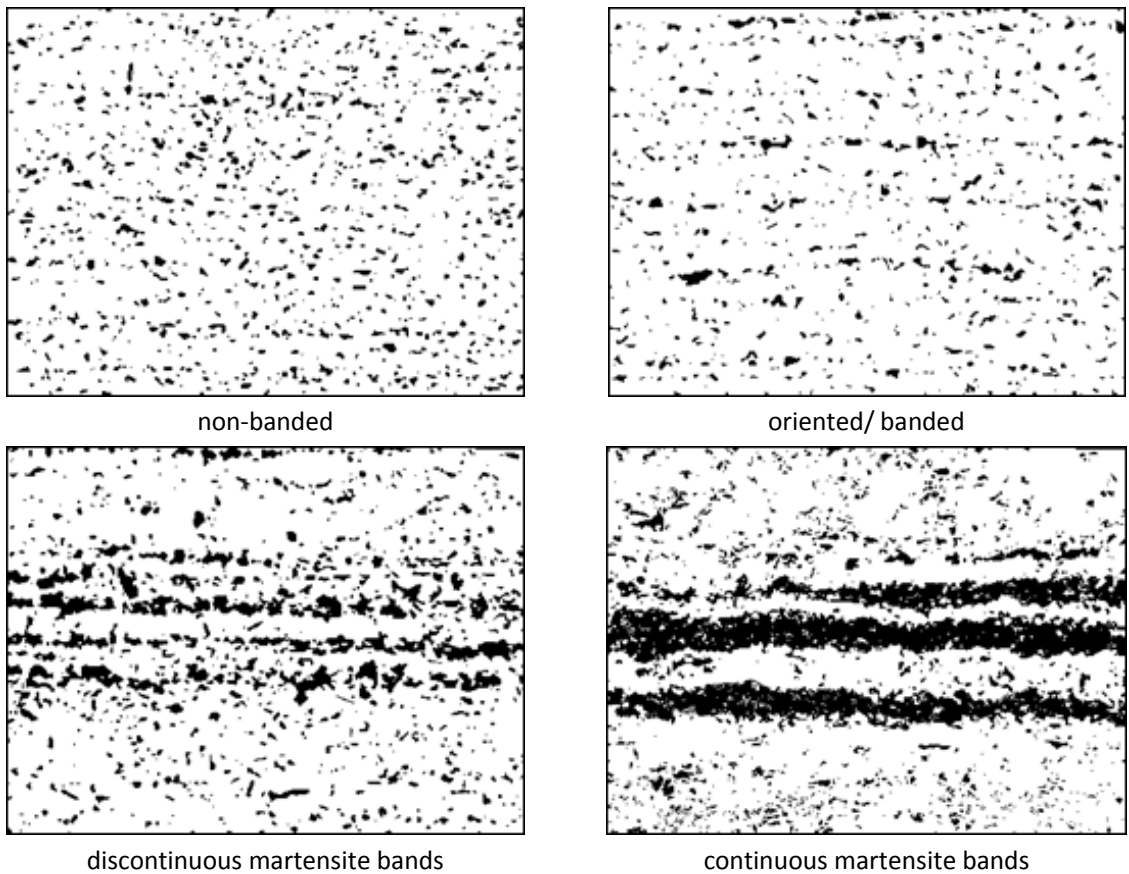

Figure 16:    Schematic representation of degrees of centre-line banding (in MP steel strip). Courtesy of Tata Steel.

Although standard line-intersection methods are available for the quantification of banding, [77] there are a number of limitations, namely the dependence upon 2D images, etching conditions, magnification, image quality and alignment. For example the aforementioned intersection standard cannot distinguish between banded and oriented samples. From and Sandström proposed a model based on local or global Fourier transforms of the grey scale in banded images [78], but complexity of the analysis and counter-intuitive nature of the results stand in the way of widespread uptake of the method.

Krebs et al proposed a covariance model based to determine quantities such as centre-to-centre distance of the bands which correspond to a wavelength, band intensity, area fraction and shape of the bands [79]. 3 parameters are derived from this method to describe banding in the microstructure: banding index (BI) to quantify intensity of phase segregation, mean distance between bands (HM) being the first maximum position of the periodicity and the slope at the origin (C'0) which is related to the fineness of the microstructure. These are illustrated in Figure 17.

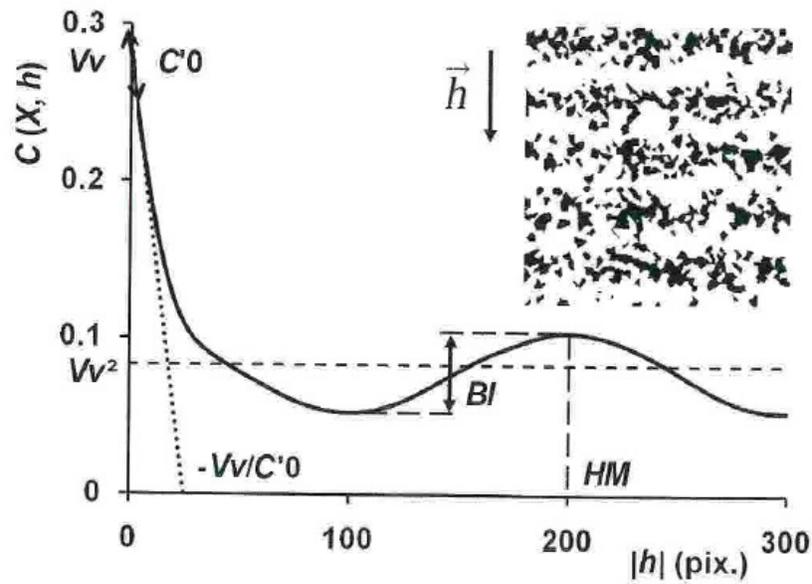

Figure 17:    Derivation of banding parameters from covariance analysis [79].

More recently McGarrity et al have extended the simple 2D descriptions banding into a 3D model using the concept of random cylinders, as shown in Figure 18 [80-83]. Although this model is outside the scope of routine image analysis, (and this literature review), it does provide an excellent input parameters for generating banded structures using 3D FEM, CA and Voronoi methods for microstructural modelling of transformations and/or mechanical behaviour.

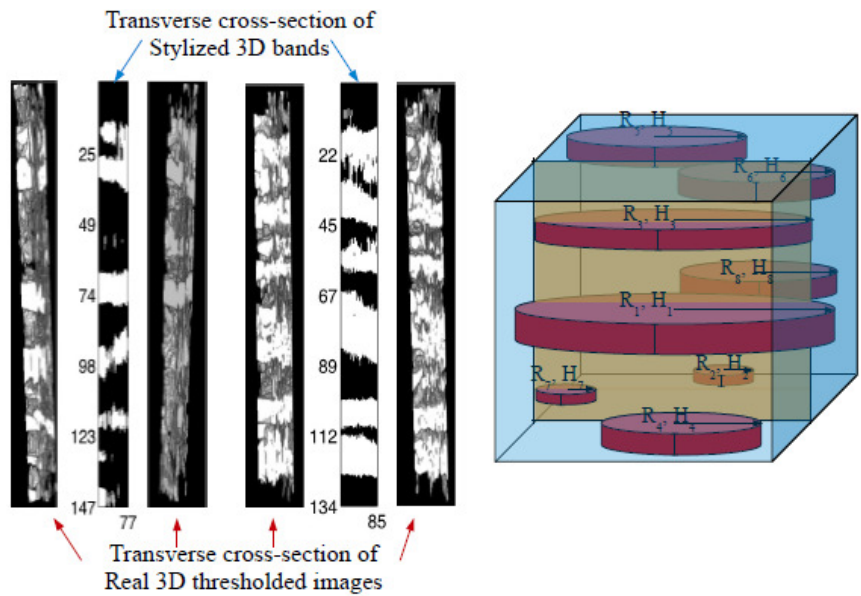

Figure 18:    Conversion of 2D to 3D quantification of banding [80-83].

## 4.1 Mechanical behaviour of ideal structures

### 4.1.1 The stress and strain tensors

For engineering purposes, transmission of forces acting on a body can be assumed to be continuous, interrupted only by structural irregularities. Analysis of the stress is based on the interaction of differently oriented surfaces passing through any point. When considering stress on a body, it is useful to consider a point dimension cube component, where the applied stress is uniform and in equilibrium and where the mass contribution can be assumed to be negligible. Each increment of the force leads to three stress components on each face of the extracted cube, leading to a total of nine stress components which are collectively known as the stress tensor [84]:

$$\begin{pmatrix} \sigma_{11} & \sigma_{12} & \sigma_{13} \\ \sigma_{21} & \sigma_{22} & \sigma_{23} \\ \sigma_{31} & \sigma_{32} & \sigma_{33} \end{pmatrix} \ or \ \begin{pmatrix} \sigma_{xx} & \tau_{xy} & \tau_{xz} \\ \tau_{yx} & \sigma_{yy} & \tau_{yz} \\ \tau_{zx} & \tau_{zy} & \sigma_{zz} \end{pmatrix} \tag{18}$$

In order for cube to be in equilibrium and not spin around the shear axes, the stress tensor must be symmetrical about the diagonal i.e. $\sigma_{ij} = \sigma_{ji}$. It follows that thus that the general stress tensor has six independent components which can be rotated into the principal stress tensor by a suitable change of axes, so that the only non-zero components of the stress tensor appear along the diagonal. In that case there are no shear stress components, only the normal stress components $\sigma_1$, $\sigma_2$ and $\sigma_3$ which are referred to as the principal stresses:

$$\begin{pmatrix} \sigma_1 & 0 & 0 \\ 0 & \sigma_2 & 0 \\ 0 & 0 & \sigma_3 \end{pmatrix} \tag{19}$$

The positions of the axes are now the principal axes and the largest principal stress is that which has a larger magnitude than any of the components found from any other orientation of the axes. The stress tensor can be separated into two stress components: hydrostatic and deviatoric, which act to change the volume or the shape of the material respectively. In crystalline metals, plastic deformation occurs by slip which conserves the material volume by changing the shape through the action of shear stresses.

### 4.1.2 Crystal plasticity

Materials can be described at several length scales from atomistic (molecular dynamics) through discrete dislocations and dislocation clusters/patterns up to grains and finally as a continuous (macroscopic) entity. When describing mechanical behaviour of crystalline materials it is not generally sufficient to consider the material only as a continuum as this discards the effects of microstructure on the properties. Behaviour at the macroscopic level, particularly plastic anisotropy is controlled by features at the microscopic level and can be described using crystal plasticity theory. The basic assumptions of this theory are as follows:

- at room temperature the major source of plastic deformation in the movement of dislocations through the crystal lattice

- dislocation movement (glide) occurs on certain crystal planes and in preferred crystallographic orientations for each type of lattice
- the crystal structure (lattice type) of metals is not altered by plastic flow
- volume changes during plastic flow are negligible

Assuming that deformation is accommodated by slip, the net result of dislocation motion is a homogenous deformation over the crystal which can be described by a linear combination of shears ($\gamma, \nabla \gamma = 0$) on the dominant slip systems. When the magnitude of the applied stress is large enough, specific slip systems are activated which consist of slip planes ($m$), being the planes of greatest atomic density, and slip directions ($s$) being close-packed direction within the slip plane. The number of slip systems is dependent upon the crystal system and the predominant slip systems for cubic lattices are shown in Figure 19.

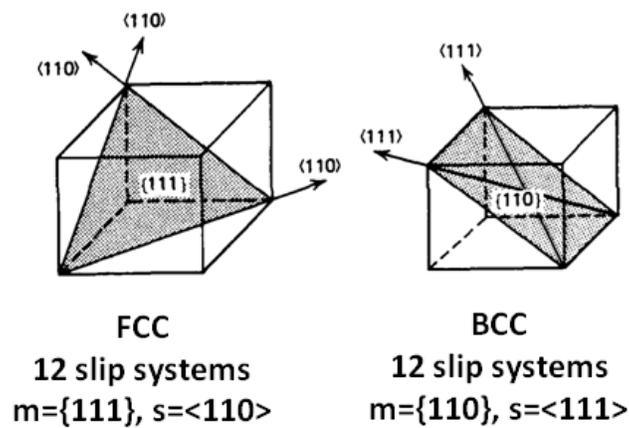

FCC
12 slip systems
m={111}, s=<110>

BCC
12 slip systems
m={110}, s=<111>

Figure 19:    Predominant slip systems in cubic lattices [85].

Schmid's law states that slip (yielding) occurs when the resolved shear stress $\tau_{rss}$ acting parallel to the slip direction on the slip plane exceeds a threshold value known as the critical resolved shear stress, $\tau_{crss}$, and is independent of the tensile stress or any other normal stress on the lattice plane [85]. The resolved shear stress can be calculated from the applied force by applying the Schmid factor, as shown graphically in Figure 20. It is apparent that the predominant slip system will be the one with the greatest Schmid factor and can be determined by calculating the Schmid factor for every slip system. The predominant slip systems for face- and body-centred cubic lattices are shown in Figure 19.

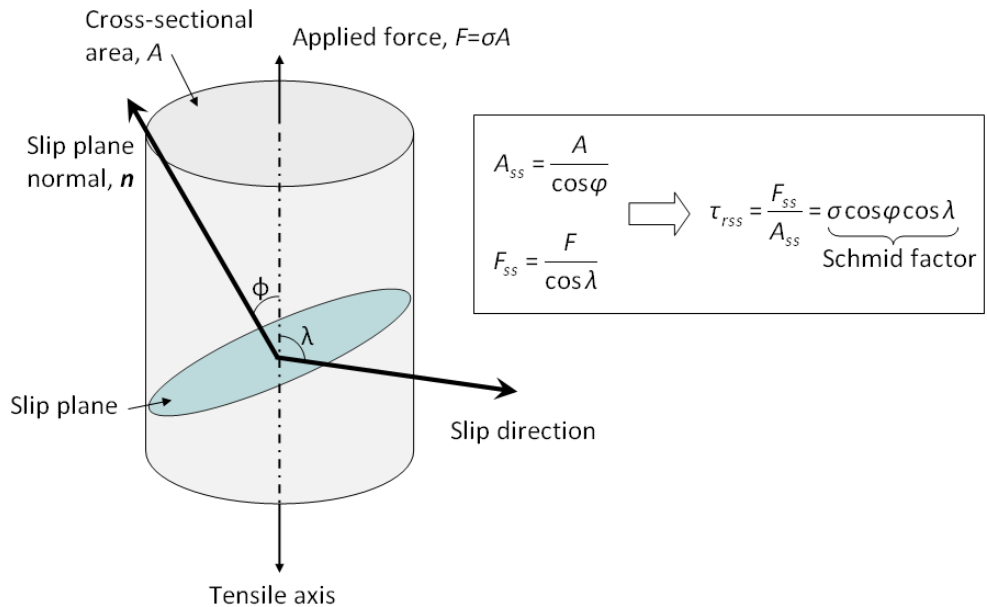

Figure 20:    Derivation of Schmid factor for a single slip system [85].

### 4.2    Mechanical behaviour in AHSS

The mechanical behaviour of DP steels has been widely studied. Generally, the mechanical properties of dual phase steels are controlled by several microstructural characteristics [86]: volume fraction of the martensite, see Figure 21 [87] with in particular the strength increasing by 10-13MPa/% martensite [88], carbon content of the martensite [89], plasticity of the martensite [20], distribution and morphology of the martensite [90] and the ferrite characteristics (grain size, carbon and alloy contents). Among these factors, the martensite fraction and carbon content are believed to play the most important role and thus have been extensively studied in the past decades.

In addition, various models considering DP [91] and TRIP [92] steels as simple composite materials have been proposed for predicting their mechanical properties, including the effect of banding [93, 94]. In fact, practical DP steels with good balance of strength and ductility have been fabricated by controlling the volume, size and morphology of two phases. However, the actual deformation of DP steels is different from a simple composite model, as ferrite and martensite must have interactions to each other during plastic deformation.

The hard phase (martensite or bainite) in DP steel has a large strain hardening coefficient, high strength and low ductility. By contrast, the soft phase (ferrite) has a low strain hardening coefficient ($n$), low yield strength ($\sigma_y$) and high ductility ($\varepsilon_U$). When the composite microstructure is stressed, the plastic strain is at first focussed in the more ductile soft phase, which work-hardens. Eventually, the harder phase also deforms plastically. This composite deformation behaviour leads to intermediate combinations of $n$, $\sigma_y$, $\varepsilon_U$. The DP microstructure thus exploits the high strength and $n$ values of the hard phase, and the high ductility of the soft phase.

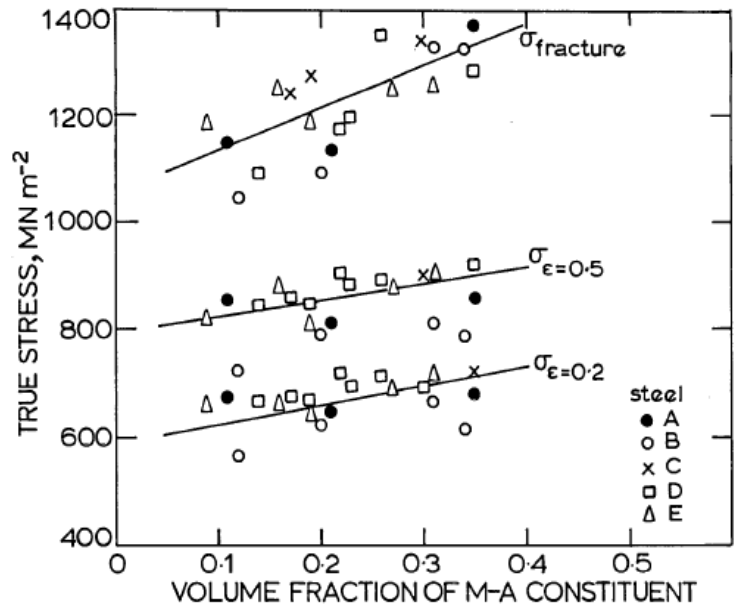

Figure 21: Effect of volume fraction of martensite and austenite on flow and fracture stresses of DP steels [87].

The fact that the plastic strain is at first focussed in the soft phase is advantageous since the hard phase can reserve its ductility until the later stages of overall deformation. The DP character may only be expected for microstructures that comprise discrete martensitic islands in a ferritic matrix. Becker et al. have shown that the critical volume fraction for interconnection of the second phase leading to decay in ductility is between 40-60% [95].

The TRIP effect leads to an increase in the uniform elongation despite the concomitant increase in yield strength. At longer BIT times, the amount and stability of retained austenite drop and this causes a respective decrease in uniform elongation. This is in agreement with Jacques [73] for Al and Si TRIP steels.

There is a direct correlation between the volume fraction of retained austenite as a function of bainite holding time and the true strain at the maximum load [96]. Vasilakos observed an increase of the yield strength with BIT time [72]. However this increase is not accompanied by a respective decrease of the uniform elongation as in other steel classes as long as carbide precipitation can be prevented. Due to the TRIP effect, the uniform elongation is maintained at high values due to the low $Ms^{\sigma}$ and high $V_f$ austenite.

### 4.3 Transformation induced plasticity (the TRIP effect)

When the austenite to martensite transformation proceeds under the influence of applied stress, it leads to a permanent plastic strain which is referred to as transformation induced plasticity or the TRIP effect. For displacive transformations such as the TRIP effect in steel, the anisotropy of each martensite variant is known and therefore the probability that a particular nucleation site for transformation would be activated can be calculated based on the lattice deformation which is the given by a combination of the invariant plane strain and the Bain distortion.

Since the displacive transformation involves the coordinated movement of atoms, it follows that the orientations of austenite and martensite lattices will be related such that when a pair of corresponding close (or densely) packed planes in the parent and daughter lattice are (nearly) parallel, so too are the corresponding directions in the respective planes. The most common orientation relationships for steel are given in Table 4.

Table 4:    Orientation relationships between parent and daughter lattices in steel

| Name | Parallel planes | Parallel directions |
|---|---|---|
| Kurdjumov-Sachs | $\{1\,1\,1\}_{\gamma} \mid\mid \{0\,1\,1\}_{\alpha'}$ | $<1\,0\,\bar{1}>_{\gamma} \mid\mid <1\,1\,\bar{1}>_{\alpha'}$ |
| Nishiyama-Wasserman | $\{1\,1\,1\}_{\gamma} \mid\mid \{0\,1\,1\}_{\alpha'}$ | $<1\,0\,\bar{1}>_{\gamma}$ ~5° from $<1\,1\,\bar{1}>_{\alpha'}$ towards $<\bar{1}\,1\,\bar{1}>_{\alpha'}$ |

The relationship between the parent fcc austenite lattice and the bct phase of the martensite resulting from transformation is shown in Figure 22, in which there is one compressive and two tensile axes. The strain required to bring about this lattice (Bain) distortion is known as the Bain strain.

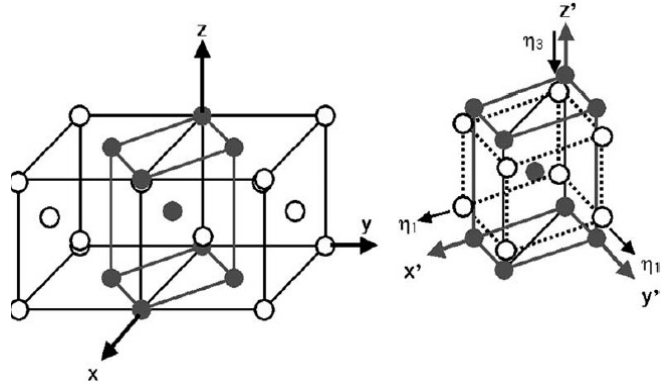

Figure 22:    Correspondence between FCC and BCT lattices showing the Bain distortion during transformation. [97]

The lattice deformation associated with the displacive transformation (*F*) without rigid body rotation can be expressed as the product of the Bain deformation and the lattice-invariant shear, respectively. The Bain deformation, *fBf*, defined on the crystallographic coordinate system of the parent austenite (FCC) can be expressed as follows:

$$fBf = \begin{bmatrix} \eta_1 & 0 & 0 \\ 0 & \eta_1 & 0 \\ 0 & 0 & \eta_3 \end{bmatrix} \tag{20}$$

Where $\eta_1$ and $\eta_3$ are the expansion and contraction ratios respectively, as shown in Figure 22, which can be calculated from the relative density ratios of the lattices. The lattice deformation associated with the transformation of austenite, *fFf*, and the resulting strain in BCT martensite, *bFb*, can be calculated using the direction and plane orientation relationships described previously.

## 4.4    Influence of mechanical loading conditions on the progression of the TRIP effect

According to the classical theory, volume expansion due to the austenite to martensite transformation results in plastic deformation and work hardening of the surrounding ferrite which in turn results in localised strengthening. This has the effect of postponing further deformation and resulting in higher stresses in the remaining untransformed austenite, which will in turn also transform thus delaying the onset of localised necking. This presumes of course that there is no grain rotation of the austenite, thereby lowering its internal stress, and the stability of very high carbon austenite.

In recent years it has been discovered that the fracture behaviour of TRIP assisted steels does not progress in accordance with the expected theory. In fact, it would appear that although the uniform elongation is increased, the available retained austenite has long since almost completely transformed to martensite [73, 74].

Tomota investigated steel with a ferrite grain size of ~2.0μm with a tensile strength of 1.1 GPa and a uniform elongation of 18.4%, in which stress-induced martensitic transformation occurred during plastic deformation [98]. A considerable amount of austenite remained even after the onset of necking and it was concluded that the enhancement of uniform elongation was caused mainly by work-hardening of hard austenite and martensite, with a negligible contribution from transformation strain. Bhadeshia [99] observed that TRIP assisted steels typically exhibit uniform tensile strains of about 15–30%; of which only about 2% may be a consequence of transformation plasticity.

This conclusion is consistent with Jacques et al [74] who argues that although low-silicon steels contain only a small amount of retained austenite, it is the composite effect that imparts superior uniform elongation. Jacques demonstrated this in Si TRIP steels with different Si and carbon contents. The calculated carbon contents were 0.95 wt% for high Si steels HSiI and HSiII and 0.73 wt% for low Si steel LSi. [74]

Figure 23 shows the amount of retained austenite present during progressive straining. The high Si steels exhibit progressive transformation with increasing strain, whilst the low Si steel transformation predominantly in the early part of plastic straining. Jacques attributes this to the carbon content of the austenite, concluding that the ductility of the low Si TRIP steel is due to a composite strengthening effect.

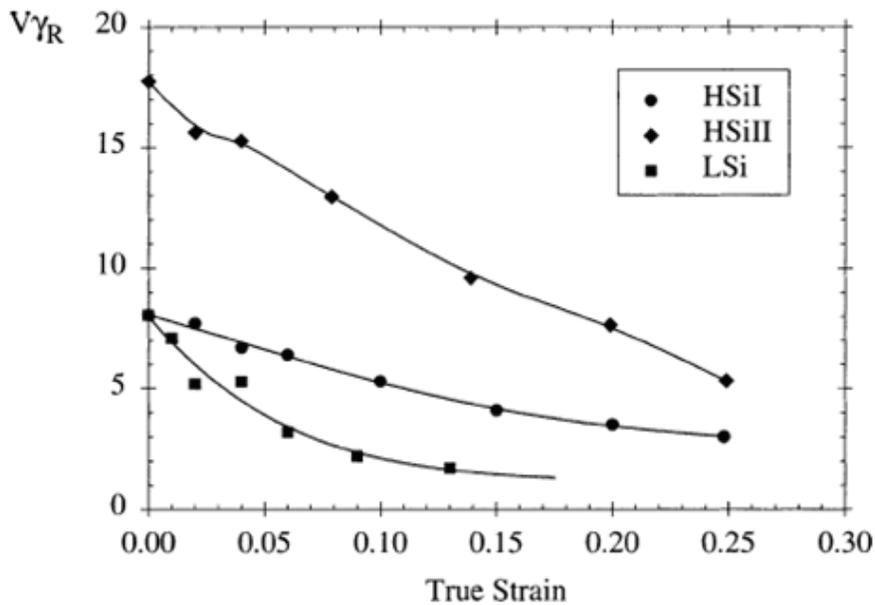

Figure 23:    Evolution of the retained austenite content during plastic straining in Si containing TRIP steels [74].

According to the rule of mixtures proposed by Olson, [100] the higher carbon martensite being be more stable would require more stress to induce transformation than the lower carbon martensite, assuming that the mechanical behaviour is consistent with stress-induced transformation (below $M_s^\sigma$). Whilst this explains the difference in transformation behaviour, it still does not explain the enhanced ductility of the low Si TRIP steel after the point at which the austenite has completely transformed.

The limitation of these measurements is that both the strains and the transformed austenite fraction are determined on bulk samples. These strains are macroscopic strains measured on an interrupted tensile test and the austenite fraction is determined by Mössbauer spectroscopy, also a bulk measurement. Furthermore, the low Si steel had a higher bulk carbon content, which for an identical austenite fraction should result in a higher carbon content in austenite, which was not the case.

In conclusion, whilst the macroscopic behaviour is determined, the progression of the transformation as a function of the local chemistry, including solid solution elements such as Mn and Al remains unexplored.

The progression of the TRIP effect has also been shown to be sensitive to test temperature, see Figure 24. Sugimoto et al [101] suggested the following relationship for the temperature and strain dependence of austenite:

$$\log(Vf_{\gamma_i}) = \log(Vf_{\gamma_0}) - k\varepsilon \qquad (21)$$

The temperature dependent factor $k$ has a minima at 100-200°C which decreases linearly with the decrease in $M_s$ for the retained austenite fraction i.e. the higher the carbon content of the retained austenite, the lower the $k$ and the more stable the retained austenite, as shown in Figure 24. This finding has been confirmed experimentally by Jimenez-Melero and co-workers [102-104].

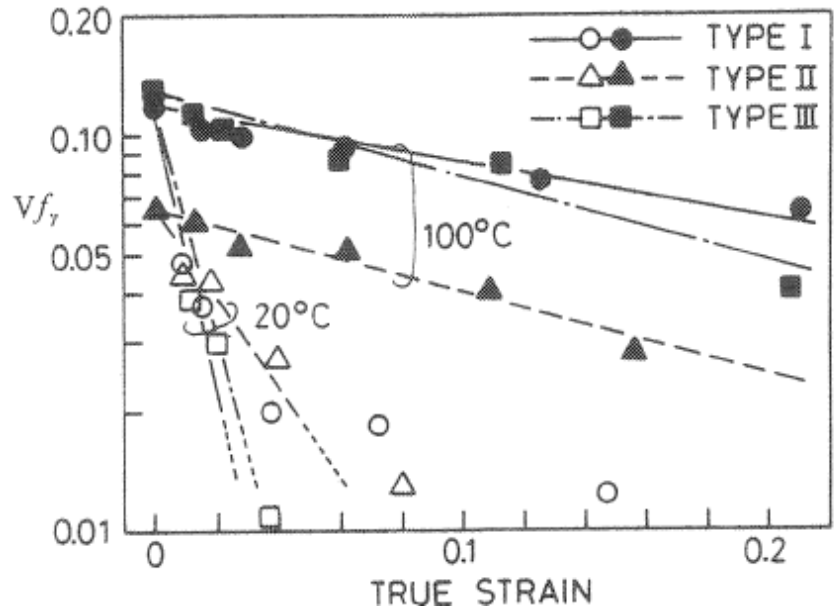

Figure 24: Effect of tensile strain and test temperature on retained austenite fraction in a TRIP-assisted dual-phase steel [90].

Takagi investigated the effect of strain rate on the TRIP effect of metastable austenitic steel and the effect of strain rate on mechanical properties are shown in Figure 25. Of particular note is the finding that noted that the volume fraction of stress-induced martensite at the same true strain decreased with an increase in strain rate [105].

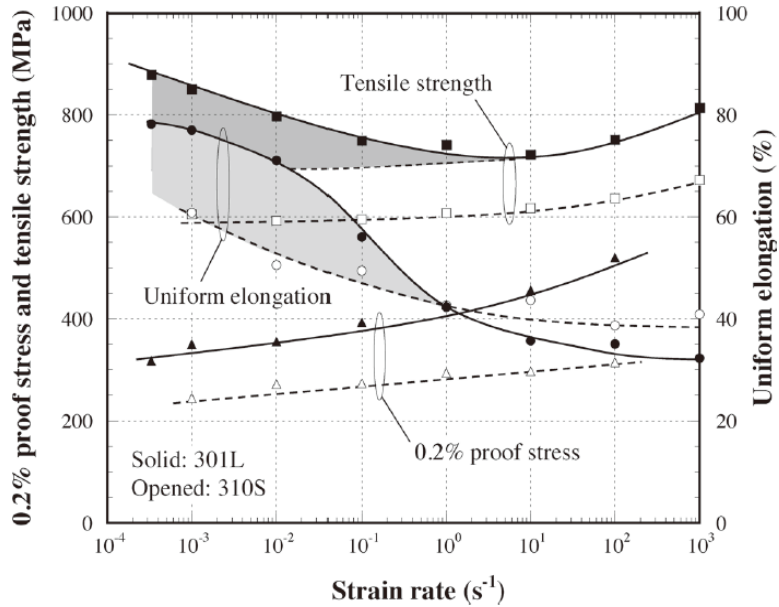

Figure 25:    Effect of strain rate on 0.2% proof stress, tensile strength and uniform elongation for two austenitic steels [105].

Yu measured the microscopic response of TRIP steels to pre-strain during plastic deformation and showed the effect of plastic strain on the dislocation density and dislocation structures which led to variation in atomic binding force and elastic modulus, see Figure 26 [106].

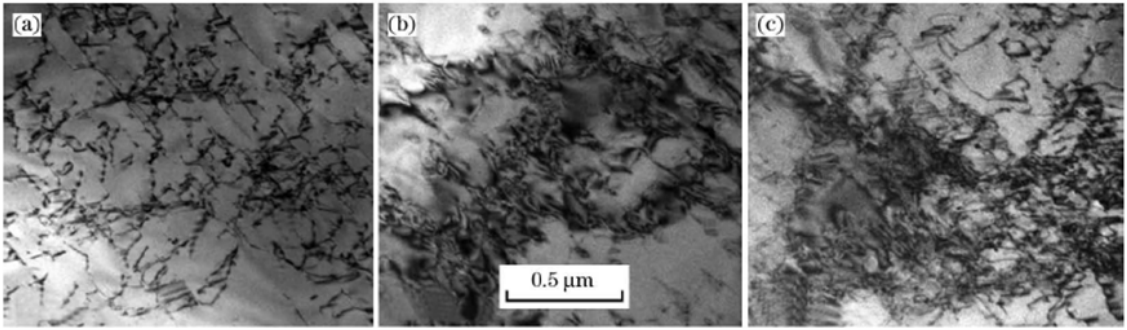

Figure 26:    Effect of plastic strain on dislocation density and dislocation structure in TRIP steel in following strain regions: (a) unstrained strain (b) pre-uniform plastic strain and (c) post-uniform plastic strain [106].

Transformation hardening in steels is also found to increase fatigue life under load-control conditions. It has also been shown that transformation can reduce crack-growth rates at low stress-intensity amplitudes , which are attributed to crack closure effects associated with the volume change resulting from transformation [100].

**4.5     Damage evolution and failure in multiphase steel**

Ductile failure is based on nucleation, growth, and coalescence of microscopic voids. According to the classical theory, nucleation occurs at inclusions and second-phase particles, of less than 1 μm, either by interfacial decohesion or by failure of the inclusions. Void growth is promoted by positive hydrostatic stress states and coalescence of voids is due either to voids accumulating in ligaments joining adjacent voids (so called 'void sheeting') or by internal necking of the ligaments. Crack advancement occurs via continuous joining of voids to the main crack. The two main factors that lead to ductile failure are thus plastic strain and hydrostatic stress

As we have already seen, multi-phase steels are designed to have a mixture of hard phases in a relatively soft matrix. This leads to a high degree of work-hardening which is beneficial for forming operations, but the difference in deformation characteristics of the phases can lead to initiation of damage. Early investigators such as Nakagawa [111] showed a positive trend between the carbon content in the martensite of DP steel with the void density, see Figure 27. This would have obvious implications for martensite resulting from transformation with carbon rich austenite.

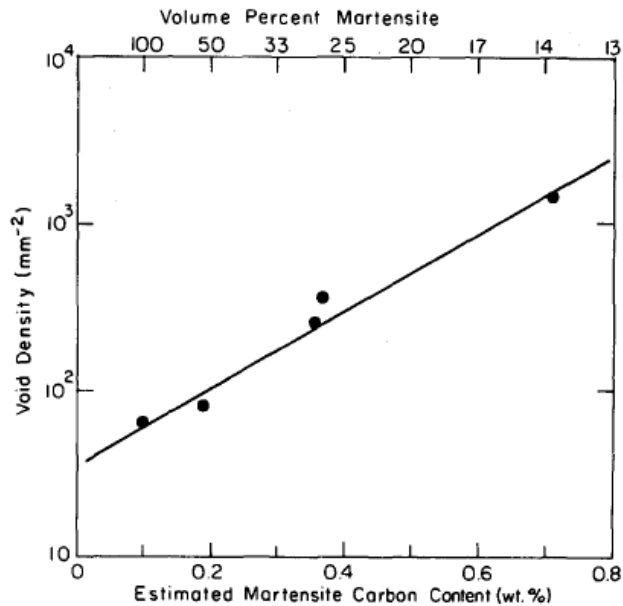

Figure 27:     Variation in void density with estimated carbon content at high strain [107].

Early work on damage in dual phase steels considered the two phases as being comparable to a composite [108, 109]. Whilst this is a useful model, the drawback is that in a composite material the phases are usually not coherent at grain boundaries which led to the speculation that the hard second phase particles either break or get de-bonded from the matrix. Whilst this is certainly true for inclusions [110], it has since been argued by Ghadbeigi and co-workers that the primary source of damage (leading to early failure) in dual phase steels is the result of strain partitioning into ferrite causing an increase in geometrically necessary dislocations (GNDs) [111]. This has recently been corroborated by Moerman et al who deduced the presence of GNDs at martensite-ferrite grain boundaries in deformed DP steel, see Figure 28 [112].

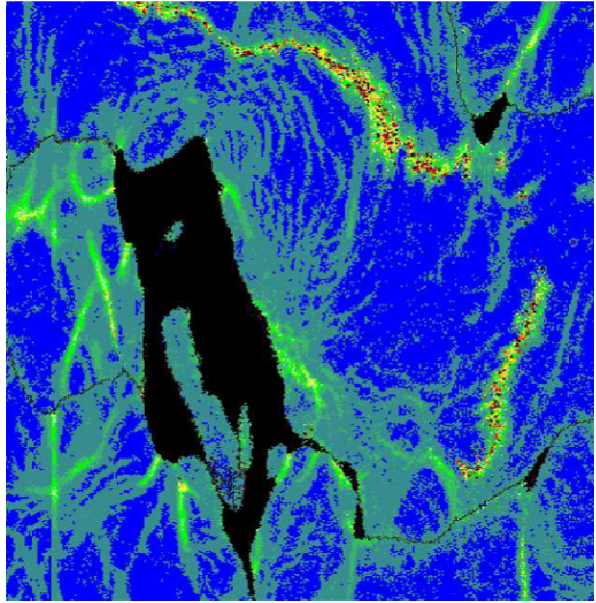

Figure 28:    EBSD map of GND sub-structure near martensite particle (black) in deformed DP steel [112].

**4.6    Effect of banding on mechanical properties**

Although it is thought that the presence of continuous bands lead to premature failure, it has not been possible to verify this link. To date, most experimental work to date looking at the effects of banding are post-mortem: the major short-coming is that it is then difficult to detect micro-voids at low strains using light microscopy. Some in-situ work has been carried out on DP steels, but this does not take into account the transformation from austenite to martensite and the effect on strain localisation due to work hardening.

Shi et al [113] investigated the role of the austenite phase distribution using finite element analysis comparing a layered (banded) austenite structure to one in which the austenite phase was randomly distributed. They observed that the highest strength was obtained in the uniformly dispersed sample and deformation was coupled with the absence of localised slip bands in the ferrite matrix, to which they attributed the strength increase. In contrast, slip bands formed within the ferrite matrix for samples where the ferrite was confined by connected austenitic grains, effectively reducing the overall strength.

The highest transformation rate was observed when the austenite layer was aligned with the loading direction. Although more transformation was accompanied with less plastic deformation (measured in terms of number of dislocations), well-defined slip bands in the ferrite matrix were observed that significantly affected the average strength. It is likely that localisation would occur in samples with dispersed transforming austenite grains at higher strain levels, however that was not investigated. Nevertheless, this work suggests that the uniform transformation of austenite during straining delays the onset of plastic localization and has higher strength than a material with a layered microstructure. This observation suggests that banded microstructures in multiphase steels assisted by transformation-induced plasticity are in general detrimental from the point of view of strength.

Based on his in situ DIC study of DP steels, Tasan [114] argues that influence of a banded structure on the global properties (YS, UTS, ductility, etc.) of DP steels is critically dependent on the morphology of the band, as well as on the mechanical behaviour of the phase that composes the band. In microstructures where there is a continuous microstructural band, shear bands are forced to develop through the band and therefore percolate through its narrowest section, see Figure 29. This forces the banded phase to deform beyond its plastic limit, especially if there is a significant difference in the ultimate strains of the phases composing the banded microstructure e.g. the case of martensite–ferrite dual phase steel.

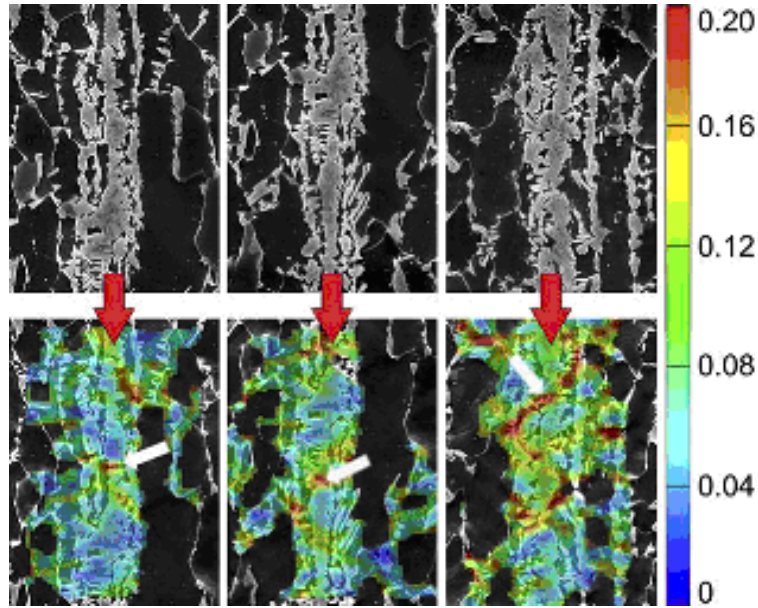

Figure 29:    Void nucleation within the band, shown here for different martensite band morphologies, for the un-deformed (above) and deformed states (below). White arrows indicate the void nucleation sites [114].

For discontinuous microstructural bands, shear bands naturally cross at the gaps within the band, thereby delaying early damage initiation. The strength of the banded phase also plays an important role; for example, a continuous band of a high fracture strength phase may accommodate high local stresses without damage initiation whereas a low fracture strength phase may not: metallographic observations demonstrated that the nucleation of micro-voids is reduced when the martensite co-deforms with the ferrite matrix [20].

Compare this last conclusion with Ghadbeigi's observation that the higher the strength of the second phase, the more strain localisation in the ferrite, causing damage accumulation. The high local strains induced between continuous bands would, therefore be more likely to cause damage in the case of high fracture strength martensite [111].

Korda et al [115] investigated the effect of ferrite-pearlite banding on fatigue crack retardation. They hypothesised that with the separation of phases into bands, the local properties (relative 'stiffness') would affect the fatigue performance: a hard phase in a soft matrix had been found to improve resistance to crack branching by deflecting the crack path. Stress intensity factor range (ΔK) decreasing/increasing and constant-ΔK fatigue crack growth tests were carried out with crack growth behaviour being observed in in situ scanning electron microscope in order to understand the influence of pearlite banding.

Fatigue crack growth resistance is ~10% higher greater in the banded orientation than in the non-banded orientation for samples with equivalent grain size. The higher threshold value was considered to arise from the barrier action of the pearlite banding. In earlier studies on fatigue crack growth in dual-phase steels, the connectivity of the hard martensite phase has also been shown to provide a higher threshold value. Hard phase connectivity is considered to offer a continuous barrier to the crack propagation. Figure 30 shows the interaction between bands and crack, with propagation taking the 'drunken man collision' route (crack branching) when crossing the path of a pearlite band. In the non-banded sample the crack

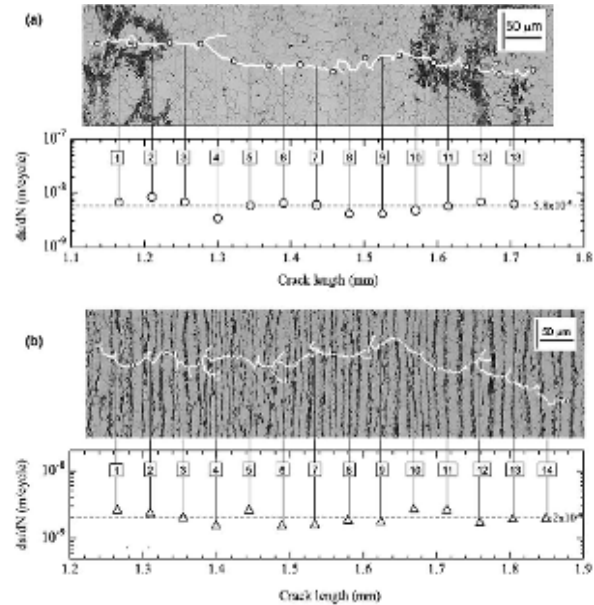

Figure 30:    Relationship between crack length and crack growth rate in constant-ΔK test for (a) non-banded (L-T) and (b) banded (L-S) orientations [115].

barely deviated from a straight line.

In conclusion, it has been shown experimentally that bands of hard martensite aligned along the loading direction can lead to increased load partitioning into the surrounding ferrite, which has been linked to the onset of early failure. The experimental work agrees with simulations of banded austenite steels and suggests that the transformation rate in aligned bands is higher than in uniformly distributes samples, but to date this remains speculation due to lack of experimental data.

### 4.7        Suppression of banding in multi-phase steels

From an industrial point of view, elimination of the banded phase is a complex task. However, even for cases where complete removal of banding is not economically feasible, the detrimental influence of a banded microstructure can be significantly reduced by altering the morphology of the band in order to (a) avoid microstructures with continuous bands and (b) decrease the thickness variation of the band. Although in principle it is possible to eliminate the Mn segregation by high temperature homogenisation treatment, as proposed by Kumar [61] due to the low diffusion rate of this element such a treatment is not practicable in the common industrial route. On the other hand, the severity of banding can be significantly reduced by selecting process settings during hot rolling and cooling on the run-out-table. In particular coarse austenite grains and high cooling rate has a beneficial effect of in reducing the severity of microstructural banding.

A predictive model quantifying the effects of thermo-mechanical treatment on band prevention is then of great use. One of these models has been developed by Offerman for hot-rolled, medium carbon steel [116]. In this model the ferrite nucleation rates are calculated in composition distinctive regions of Mn, Si and Cr; and based on experimental data, it is postulated that when the difference in ferrite nucleation rates exceeds 6–8%, ferrite/pearlite banding occurs.

Xu et al further developed this concept by combining it with the calculation of segregation due to solidification employing a thermo-chemical database combined with diffusion theory and classical nucleation theory [117]. The influence of processing parameters in banding was calculated for a number of industrial steel grades by introducing "band prevention plots" as a means to relate the austenitisation and overaging temperatures with the austenitisation time and micro-chemical wavelength. Their results were consistent with the experimentally observed conditions for ferrite/pearlite band prevention reported for a number of steel grades in the literature. It was reported that ferrite/pearlite banding disappears when the austenite grain size exceeds the chemical banding wavelength by a factor 2 or 3, since the influence of grain boundaries as a preferred sites for nucleation dominates the effect of the compositional gradient.

For the critical case of reducing the severity of banding in DP steels, for example, this may be achieved by optimising the critical production parameters (e.g. rate of cooling during hot rolling, intercritical annealing temperature, soaking duration) that have a significant influence on the degree of banding. The work of Mecozzi [118, 119] on the cellular automata simulation (CAS) programme has shown that the chemical segregation of a number of elements present in AHSS (namely Mn, Al, Cr) are important pre-cursors to banding after hot-rolling. With regard to the effect of cooling rate, the difference of the start temperature for ferrite nucleation in high and low Mn regions is reduced by increasing the cooling rate and therefore a more homogeneous microstructure is created upon cooling. The preliminary model results show a good agreement with experimental results for banded structures in the hot-rolled condition based on line scan measurements of AHSS.

Cold-rolling and annealing is necessary to achieve the thicknesses and mechanical properties required for most automotive products. During the annealing of cold-rolled material, recovery and recrystallisation occur at lower temperatures than those required for austenite transformation. Furthermore, the aforementioned alloying elements affect the austenite transformation temperature on heating and cooling.

As has already been demonstrated, Mn is the alloy element that is primarily responsible for the development of microstructural banding in steels due to the effect of this element on the ferrite/austenite equilibrium temperature $A_{e3}$. Ferrite starts to form during cooling in bands with low Mn content (high $A_{e3}$) because the local undercooling favours nucleation. The formation of ferrite is accompanied by the rejection of C into the adjacent high Mn bands; these C and Mn rich austenite bands later transform to pearlite resulting in a banded ferrite and pearlite microstructure.

Caballero et al [120] investigated the relationship between chemical segregation, hot-strip mill processing and annealing parameters on the incidence of microstructural banding. They found that the segregation of manganese during solidification from casting is responsible for banding problems of dual phase steel. The study revealed although that banding increasing the cooling rate during hot rolling does suppress the formation of ferrite-pearlite banding, this is only true of the intermediate microstructure. Upon intercritical annealing at high temperatures (in this case 800°C), the degree of banding increases as the transformation proceeds until finally it resembles the original chemical segregation. However, it was shown that permanently eliminating microstructural banding was possible by annealing low in the intercritical range (750°C) with a longer soaking time (100 s). It is therefore evident that although the effects of banding brought about by chemical segregation during casting can be suppressed by careful selection of hot-rolling parameters, this can be undone during annealing.

## 5.    In-situ characterisation techniques

This chapter explores the theory and application of the experimental techniques for in situ characterisation which will be employed in this work.

### 5.1    X-ray diffraction

As previously elucidated in section 4.4, much experimental work has been carried out in attempts to determine the effect of mechanical loading on the progression of the austenite to martensite transformation. However, since the transformation is influenced by microscopic features, such as grain size and morphology, the local stress state and, in the case of banding orientation to the stress axis and the nature of the surrounding matrix, has made macroscopic analysis of this phenomenon extremely difficult for fine grained, multiphase steels.

More recently advances have been made using synchrotron radiation to monitor the transformation at the level of individual grains, allowing quantification of the effect of local carbon content and stress on thermal and mechanical stability of austenite [102, 104, 121, 122]. The availability of high intensity X-ray beams in energy ranges of $50-300$ keV, combined with micro-focussing optics giving spot sizes at the sample in the micron range without loss of flux, has led to a revolution in the science of X-ray diffraction of metallic alloys.

X-ray diffraction in this work is based on the principle of elastic (Thompson) scattering such that the wavelength and energy of the deflected X-rays are not changed, only the momentum. Diffracted waves from different atoms interfere with each other and the resultant intensity distribution is strongly modulated by this interaction. If the atoms are arranged in a periodic fashion, as in crystals, the diffracted waves will consist of sharp interference maxima (peaks) with the same symmetry as in the distribution of atoms. The relationship between the spatial distribution of parallel planes and the angle and wavelength of the incident beam are described by the Bragg Law:

$$n\lambda = 2d\sin\theta \qquad\qquad (22)$$

Where $n$ is an integer multiplier of the inter-planar distance,[c] $\lambda$ is the wavelength of incident wave (Å), $d$ is the spacing between the planes in the atomic lattice (Å), and $\theta$ is the angle between the incident ray and the scattering planes. The inter-planar spacing, $d$, can be calculated from the Miller indices *(h k l)* of the lattice. For the cubic system the equation simplifies due to the symmetry of the unit cell, thus:

$$d_{(hkl)} = \frac{a}{\sqrt{h^2 + k^2 + l^2}} \quad\quad\quad (23)$$

---

[c] The integer $n$ refers to the order of diffraction. For $n$=1, $d\sin\theta$ =1 and for $n$=2, $d\sin\theta$ = 2, e.g. for planes related to (112) and separated by 4 Å: $d$(112) = 4 Å, $d$(224) = 2 Å, $d$(336) = 1.333˙ Å, etc.

By modelling a crystal as a set of discrete parallel planes separated by a constant parameter $d$, it can be deduced that the incident X-ray radiation would produce a Bragg peak if their reflections off the various planes interfered constructively, which occurs at a phase shift of multiples of 2π. By changing the geometry of the incident rays, the orientation of the centred crystal and the detector, it is possible to attain all possible diffraction directions of the lattice.

The concept of Bragg diffraction applies equally to neutron diffraction and electron diffraction processes. Both neutron and X-ray wavelengths are comparable with inter-atomic distances (~150 pm) and thus are an excellent probe for this length scale. Since X-rays are directed at the sample, and the diffracted rays are collected, a key component of all diffraction is the angle between the incident and diffracted rays. Powder and single-crystal diffraction vary in instrumentation beyond this.

## 5.2    Powder diffraction

Powder X-ray diffraction (pXRD) is perhaps the most widely used x-ray diffraction technique for characterizing crystalline materials. As the name suggests, the sample is usually in a powdery form, consisting of fine grains of single crystalline material to be studied. The technique is used also widely for studying particles in liquid suspensions or polycrystalline solids (bulk or thin film materials).

The term 'powder' really means that the crystalline domains are randomly oriented in the sample. Therefore when the 2-D diffraction pattern is recorded, it shows concentric rings of scattering peaks corresponding to the various $d$ spacings in the crystal lattice. By integrating the diffraction pattern, a line spectrum is obtained from which structural information is obtained by least squares fitting of the peaks to known crystallographic parameters of the phases present, the most widely used being the Rietveld method.

The positions and the intensities of the peaks are used for identifying phases and relative phase fractions, the shape of the peak can be used to derive the lattice. Powder diffraction patterns can be collected using either transmission or reflection geometry, as shown in Figure 31. Because the particles in the powder sample are randomly oriented, these two methods will yield the same data; the major difference being that the reflection geometry yields information about the surface geometry and transmission, the bulk. Typical applications of the reflection geometry are the analysis of dendrite growth during welding operations [123] and oxide growth [124]. When combined with chemical analysis techniques, such as X-ray fluorescence (XRF), this can be a powerful tool for analysing the role of alloying elements on the thermal and, in our case, mechanical behaviour. The geometry outlined by Bischoff *et al.* [124] was used as a basis for one of the experiments in this work.

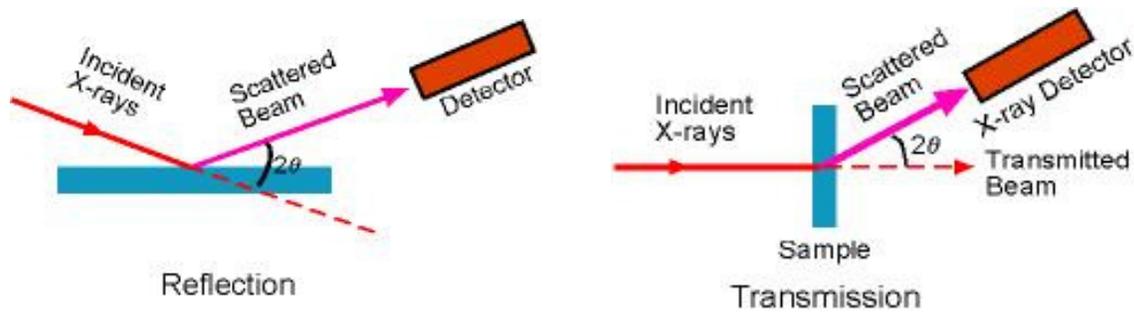

Figure 31: Illustration of the difference between reflection and transmission geometry in XRD experiments

The limitation of this method is that the difficulty in building up statistically significant data. Furthermore, one has to take into account role of the free surface in inducing the austenite to martensite transformation and in particular stresses induced due to preparation. In such cases care must be taken to avoid inducing stresses or, failing that, removing deformed surface layers e.g. by electropolishing.

In transmission mode, information can be obtained over much larger volumes, yielding more statistically more reliable information. By rotating the sample around its axis, as illustrated in Figure 32 it is possible to elucidate three-dimensional microstructural information down to the level of individual grains. Since the majority of the diffraction pattern originates within the bulk of the material, surface effects are generally negligible.

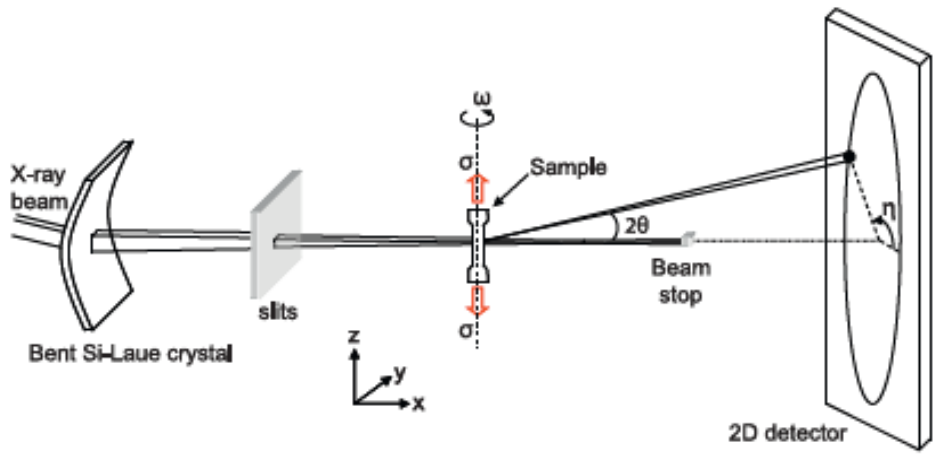

Figure 32: Schematic representation of the HEXRD setup used for the in-situ synchrotron diffraction experiments under uniaxial tensile stress [104].

The beam size is defined by the slits situated between the silicon monochromator and the sample. The Si-Laue crystal scatters vertically and is bent towards the vertical around a horizontal axis. The tensile rig is placed on a table that can be translated in three dimensions and rotated along the cylindrical axis of the sample ($\omega$-rotation). The load ($\sigma$) is applied perpendicular to the incoming X-ray beam. The diffracted intensity is collected on a two-dimensional CCD detector placed behind the sample upon which the scattering angle ($2\theta$) and the azimuth angle ($\eta$) are indicated in the figure.

**5.3    Application to mechanical behaviour**

Evolution of intensity of *{h k l}* reflections coming from each phase during deformation provides information about changes in phase fraction and texture in response to the applied load, see Figure 33 in which it can be seen from the diffraction pattern that the intensity of the austenite $(220)_\gamma$, $(311)_\gamma$ and $(200)_\gamma$ reflections are reduced after 11% strain indicating mechanically induced transformation to martensite.

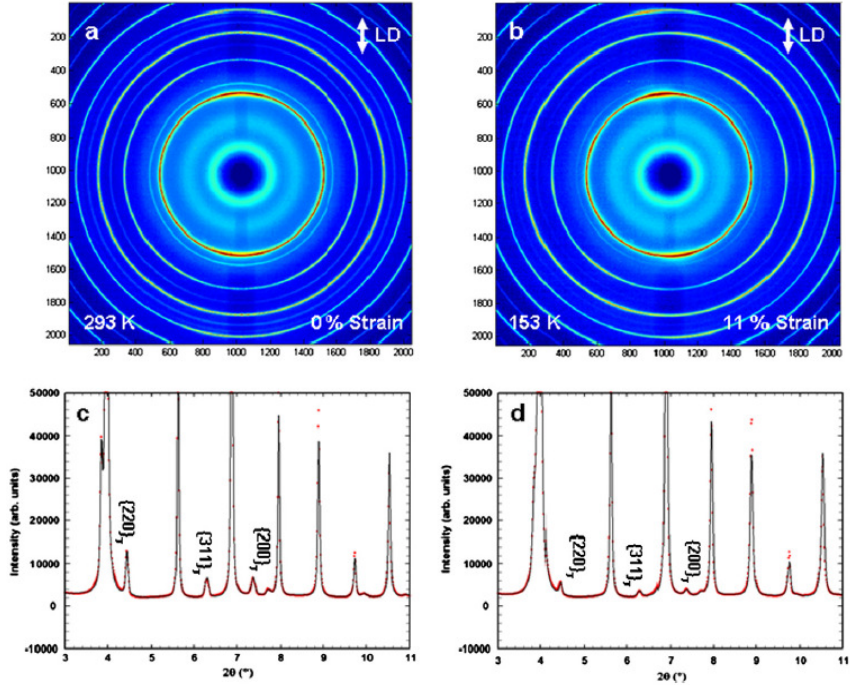

Figure 33:    Effect of strain on the evolution of austenite phase fraction in a low alloyed TRIP steel [125].

Shifts in peak positions reveal changes in the average lattice parameters from the original condition. The peak shifts can be translated into the corresponding average elastic phase strains $\left( \left\langle \varepsilon_{ph} \right\rangle \right)$ as follows:

$$\left\langle \varepsilon_{ph} \right\rangle = \frac{a_{ph}^i - a_{ph}^0}{a_{ph}^0} \qquad (24)$$

Where $a_{ph}^0$ and $a_{ph}^i$ are the original lattice parameter and the corresponding lattice parameter at the *i*th strain step. The evolution of $\left\langle \varepsilon_{ph} \right\rangle$ with the applied stress provides information about the load partitioning between different phases. This technique can be further confined to different grains within each phase which satisfy the Bragg condition for specific *{h k l}* planes and grain orientations with respect to the incident beam. In this case the observed shift in peak positions is manifested by an increase in the average lattice plain strain $\left\langle \varepsilon_{hkl} \right\rangle$ which is calculated using the following expression:

$$\left\langle \varepsilon_{hkl} \right\rangle = \frac{d_{hkl}^i - d_{hkl}^0}{d_{hkl}^0} \qquad (25)$$

Where $d_{hkl}^0$ and $d_{hkl}^i$ are the original lattice plane spacing and the corresponding lattice plane spacing at the $i$th strain step. In this way it is possible to track the inter- and intra-phase load partitioning on the interplay thereof with mechanically induced austenite to martensite transformation and mechanical behaviour.

### 5.4    Digital image correlation

Digital image correlation is an optical method to measure displacement of local points by comparing digital images of the un-deformed and deformed specimen, as illustrated with a unit grid in Figure 34 [126].

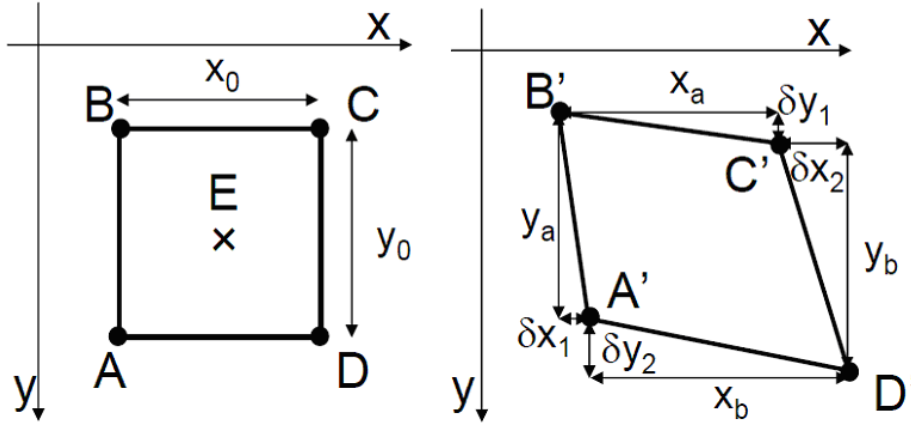

Figure 34:    Calculation of local strain based on displacement between original and strained geometry [126].

In a real microstructure, grid points can either be superimposed on the surface prior to tensile testing, or in the case of microstructures with very distinctive feature differences e.g. DP steels, the microstructure itself can be used to track displacement and thence, strain.

Local strains are calculated from the distance between neighbouring points. Firstly, the four corners of the unit grid in the un-deformed image are denoted A, B, C and D. After one deformation step in which the stress is applied in the x direction, these points are displaced to A', B', C' and D', respectively. The strains parallel to the x and y directions, $\varepsilon_{xx}$ and $\varepsilon_{yy}$ respectively, the shear strain, $\varepsilon_{xy}$ and the equivalent strain, $\varepsilon_{eq}$ are calculated as follows:

$$\varepsilon_{xx} = \frac{1}{2}\left( \frac{x_a - x_0}{x_0} - \frac{x_b - x_0}{x_0} \right) \tag{26}$$

$$\varepsilon_{yy} = \frac{1}{2}\left( \frac{y_a - y_0}{y_0} - \frac{y_b - y_0}{y_0} \right) \tag{27}$$

$$\varepsilon_{xy} = \frac{1}{2}\left\{ \left( \frac{\delta x_1}{y_a} + \frac{\delta y_1}{x_a} \right) + \left( \frac{\delta x_2}{y_b} + \frac{\delta y_2}{x_b} \right) \right\} \tag{28}$$

$$\varepsilon_{eq} = \sqrt{\left\{ \frac{2}{3}\left(\varepsilon_{xx}^2 + \varepsilon_{yy}^2\right) + \frac{1}{3}\left(\varepsilon_{xy}^2\right) \right\}} \qquad (29)$$

## 5.5    Tomography

Since it is the intention to carry out an experiment using 3D tomography to determine the interaction between the TRIP effect and the onset of damage under loading, a brief outline of tomography is appropriate at this point. Tomography is science of combining multiple projections of an illuminated body to construct a three-dimensional image of its structure. This is an extension into 3D of traditional X-ray photography as used in medical examination and is heavily dependent on absorption contrast within the material.

According to the Beer–Lambert law, the ratio of the number of transmitted to incident photons is related to the integral of the absorption coefficient of the material $\mu$ along the path that the photons follow through the sample. When the beam is monochromatic, the absorption coefficient can be related empirically to density, atomic number and energy. The resulting image is a superimposed projection of a volume in a 2D plane [127]. Three different modes are available to perform tomography:

**Absorption mode**: whereby contrast is given by the difference between the linear attenuation coefficients $\mu_1$ and $\mu_2$, which are in turn directly related to atomic number and density, thereby facilitating semi-quantitative analysis through differences in grey scale level.

**Phase contrast mode**: by increasing the distance between the sample and the detector it is possible to induce phase interference which with superposition on the original (absorption) contrast, is efficient for edge detection.

**Holo-tomography mode**: this mode uses images taken at several sample to detector distances whereby the quantitative distribution of the optical phase can be retrieved. Using tomographic reconstruction it is possible to specific refractive index for each voxel, which is proportional to the electron density of the material and allows 3D mass density quantification.

## 6.    References

[1]   P. Mock, EU CO2 standards for passenger cars and light-commercial vehicles, International Council on Clean Transportation, http://www.theicct.org/eu-co2-standards-passenger-cars-and-lcvs

[2]   International Council on Clean Transportation, Global passenger vehicle data sheet, http://www.theicct.org/sites/default/files/info-tools/Global_PV_Std_Aug2013_data.xlsx

[3]   O. Hoffmann, Lightweight steel design in the modern vehicle body, in: Werkstoff-Forum Intelligenter Leichtbau, Hannover, Germany, (2011)

[4]   European Parliament, European Council, Regulation (EC) No 443/2009 of the European Parliament and of the Council of 23 April 2009 setting emission performance standards for new passenger cars as part of the Community's integrated approach to reduce CO 2 emissions from light-duty vehicles, Regulation (EC) No 443/2009 (2009)

[5]   S.M. Zoepf, Automotive features - mass impact and deployment characterisation, MSc Thesis, Massachusetts Institute of Technology, Cambridge, MA, USA, 2011

[6] H. Hofmann, D. Mattissen, T.W. Schaumann, Advanced cold rolled steels for automotive applications, Steel Research Int., 80 (2009) 22-28. doi: 10.1002/srin.200990094

[7] M. Takahashi, Development of high strength steels for automobiles, Nippon Steel Technical Report, 88 (2003) 2-7

[8] J. Zrník, I. Mamuzic, S.V. Dobatkin, Recent progress in high strength low carbon steels, Metallurgija, 45 (2006) 323-331

[9] K. Tanaka, T. Saito, Phase equilibria in TiB2-reinforced high modulus steel, Journal of Phase Equilibria, 20 (1999) 207-214

[10] N. Sugiura, N. Yoshinaga, S. Hiwatashi, M. Takahashi, K. Hanya, N. Uno, R. Kanno, A. Miyasaka, T. Senuma, Steel sheet having high Young's modulus, hot-dip galvanized steel sheet using the same, alloyed hot-dip galvanized steel sheet, steel pipe having high Young's modulus and methods for manufacturing the same, Nippon Steel Corporation, Patent No. US8057913, USA, 2005

[11] T. Gréday, H. Mathy, P. Messien, About different ways to obtain multiphase steels, in: R.A. Kot (Editor), Structure and properties of dual-phase steels, Chicago, IL, USA, (1979)

[12] R.A. Kot, B.L. Bramfitt, Fundamentals of dual phase steels: Conference Proceedings, in: R.A. Kot and B.L. Bramfitt (Eds.), Fundamentals of Dual-Phase Steels, Chicago, IL, USA, (1981)

[13] G.A. Kunitsyn, S.V. Denisov, A.V. Gorbunov, A.G. Vetrenko, A.I. Brus'yanina, E.V. Zharkov, Production of high-strength steel sheet for the auto industry, Steel in translation, 38 (2008) 585-588

[14] B. Ehrhardt, T. Gerber, T.W. Schaumann, Approaches to microstructural design of TRIP and TRIP aided cold rolled high strength steels, in: International Conference on Advanced High Strength Sheet Steels for Automotive Applications, Winter Park, CO, USA, (2004)

[15] R. Mostert, B.L. Ennis, D.N. Hanlon, Adapting AHSS concepts to industrial practice, in: International Symposium on New developments in Advanced High-Strength Sheet steel, Vail, CO, USA, (2013)

[16] D.K. Matlock, J.G. Speer, Third generation of AHSS - microstructure design concepts, in: A. Haldar, S. Suwas and D. Bhattacharjee (Eds.), Microstructure and texture in steels and other materials, Jamshedpur, India, (2008)

[17] B.L. Ennis, D.N. Hanlon, High strength hot-dip galvanised steel strip, Tata Steel IJmuiden B.V., Patent No. WO2011076383, 2010

[18] World Auto Steel, Types of steel used in the automotive industry, http://www.worldautosteel.org/steel-basics/steel-types/

[19] S. Oliver, T.B. Jones, G. Fourlais, Dual phase versus TRIP strip steels: Microstructural changes as a consequence of quasi-static and dynamic tensile testing, Mater. Charact., 58 (2007) 390-400

[20] M. Mazinani, W.J. Poole, Effect of martensite plasticity on the deformation behavior of a low carbon dual-phase steel, Metall. Mater. Trans. A, 38A (2007) 328-339

[21] W. Bleck, Using the TRIP effect - the dawn of a promising group of cold formable steels, in: International Conference on TRIP-aided High Strength Ferrous Alloys, Gent, Belgium, (2002)

[22] P.J. van der Wolk, Modelling CCT diagrams of engineering steels using neural networks, PhD Thesis, Delft University of Technology, The Netherlands, 2001

[23] S. Kim, C. Lee, T. Lee, C. Oh, Effect of Cu, Cr and Ni on mechanical properties of 0.15wt.%C TRIP-aided cold rolled steels, Scripta Mater., 48 (2003) 539-544

[24] F.D. Fischer, G. Reisner, E. Werner, K. Tanaka, G. Cailletaud, T. Antretter, A new view on transformation induced plasticity (TRIP), Int. J. Plasticity, 16 (2000) 723-748. doi: 10.1016/S0749-6419(99)00078-9

[25] M. Gomez, C.I. Garcia, D.M. Haezebrouck, A.J. DeArdo, Design of composition in (Al/Si)-alloyed TRIP Steels, ISIJ Intl., 49 (2009) 302-311

[26] H.K.D.H. Bhadeshia, D.V. Edmonds, The bainite transformation in a silicon steel, Metall. Trans. A, 10A (1979) 895-907

[27] R. Bode, M. Meurer, T.W. Schaumann, W. Warnecke, Selection and use of coated AHSS for automotive applications, Stahl Eisen, 8 (2004) 19-24

[28] K. Eberle, P. Cantinieaux, P. Harlet, New thermomechanical strategies for the production of high strength low alloyed multiphase steel showing a transformation induced plasticity (TRIP) effect, Steel Res., 70 (1999) 233-238

[29] J. Maki, J. Mahieu, B.C. De Cooman, S. Claessens, Hot-dip galvanising of Si-free C-Mn-Al TRIP steels, in: Marcel Lamberights and Centre de Recherches Metallurgiques (CRM) (Eds.), Galvatech 2001, Brussels, Belgium, (2001)

[30] M. De Meyer, D. Vanderschueren, B.C. De Cooman, The influence of the substitution of Si by Al on the properties of cold-rolled C-Mn-Si TRIP steels, ISIJ Intl., 39 (1999) 813-822

[31] T. Gladman, Solubility of microalloy carbides and nitrides, in: The physical metallurgy of microalloyed steels, Maney Publishing, London, UK, 2002. ISBN 978 1 902653 81 5

[32] A. Lucas, J.C. Herman, A. Schmitz, Cu-containing TRIP steels, in: International Conference on TRIP-aided High Strength Ferrous Alloys, Gent, Belgium, (2002)

[33] C.K. Ande, M.H.F. Sluiter, First-principles prediction of partitioning of alloying elements between cementite and ferrite, Acta Mater., 58 (2010) 6276-6281. doi: 10.1016/j.actamat.2010.07.049

[34] H.C. Chen, H. Era, M. Shimizu, Effect of P on the formation of retained austenite and mechanical properties in Si-containing LC steel sheet, Metall. Trans. A, 20A (1989) 437-445

[35] A. Pichler, P. Stiaszny, TRIP steel with reduced silicon content, Steel Res., 70 (1999) 459-465

[36] S. Traint, A. Pichler, K. Hauzenberger, P. Stiaszny, E. Werner, Influence of silicon aluminium phosphorus and copper on the phase transformations of low alloyed TRIP-steels, in: International Conference on TRIP-aided High Strength Ferrous Alloys, Gent, Belgium, (2002)

[37] L. Barbé, L. Tosal-Martínez, B.C. De Cooman, Effect of phosphorus on the properties of a cold rolled and intercritically annealed TRIP-aided steel, in: International Conference on TRIP-aided High Strength Ferrous Alloys, Gent, Belgium, (2002)

[38] M. Amirthalingam, N. den Uijl, M.J.M. Hermans, Fosfor- en boriumsegregatie tijdens weerstandpuntlassen van geavanceerde hogesterktestalen, Lastechniek, 78 (2012) 18-21

[39] F.B. Pickering, Structure-property relationships in steels, in: Cahn et al. (Eds.), Materials Science and Engineering, Wiley, 1993, pp. 45-79

[40] W. Bleck, A. Frehn, J. Ohlert, Niobium in dual phase and TRIP steels, in: International Symposium on Niobium 2001, Orlando, FL, USA, (2011)

[41] A.Z. Hanzaki, P.D. Hodgson, S. Yue, The influence of bainite on retained austenite characteristics in Si-Mn TRIP steels, ISIJ Intl., 35 (1995) 79-85

[42] K. Sugimoto, T. Murumatsu, S. Hashimoto, Y. Mukai, Formability of Nb bearing ultra-high strength TRIP-aided sheet steels, J. Mat. Proc. Tech., 177 (2006) 390-395

[43] Wikipedia, Solidification schematic used for deriving the Scheil-Gulliver equation, http://en.wikipedia.org/wiki/File:Scheil_graphics.gif

[44] M.C. Flemings, Solidification Processing, McGraw-Hill, Inc., New York, 1974

[45] M. Xiong, A.V. Kuznetsov, Comparison between Lever and Scheil rules for modeling of microporosity formation during solidification, Flow. Turbul. Combust., 67 (2001) 305-323. doi: 10.1023/A:1015291706970

[46] G. Krauss, Solidification, segregation and banding in carbon and alloy steels, Metall. Mater. Trans. B, 34B (2003) 781-792

[47] R.M. Fisher, G.R. Speich, L.J. Cuddy, H. Hu, Phase transformations during steel production, in: Proceedings of Darken Conference "Physical Chemistry in Metallurgy", Monroeville, PA, USA, (1976)

[48] D. Chakrabarti, C.L. Davis, M. Strangwood, Development of bimodal grain structures in Nb-containing High-Strength Low-Alloy steels during slab reheating, Metall. Mater. Trans. A, 39 (2008) 1963-1977

[49] J. Black, Modeling the effects of chemical segregation on phase transformations in medium carbon bar steels, MSc Thesis, Colorado School of Mines, Golden, CO, USA, 1998

[50] E.J. Pickering, Macrosegregation in steel ingots: The applicability of modelling and characterisation techniques, ISIJ Intl., 53 (2013) 935-949

[51] W.R. Irving, Continuous casting of steel, The Institute of Materials, London, UK, 1993

[52] E.J. Pickering, M. Holland, Detection of macrosegregation in a large metallic specimen using XRF, Ironmaking & Steelmaking, (2014) . doi: doi: 10.1179/1743281213Y.0000000144

[53] H. Mizukami, K. Hayashi, M. Numata, A. Yamanaka, Prediction of solid-liquid interfacial energy of steel during solidification and control of dendrite arm spacing, ISIJ Intl., 52 (2012) 2235-2244

[54] M.H. Burden, J.D. Hunt, Cellular and dendritic growth, J. Cryst. Growth, 22 (1973) 109-116. doi: http://dx.doi.org/10.1016/0022-0248(74)90127-4

[55] W. Kurz, B. Giovanola, R. Trivedi, Theory of microstructural development during rapid solidification, Acta Metall., 34 (1986) 823-830. doi: http://dx.doi.org/10.1016/0001-6160(86)90056-8

[56] L. Tan, Multi-scale modelling of solidification of multi-component alloys, PhD Thesis, Cornell University, Ithaca, NY, USA, 2007

[57] P. Wycliffe, Microanalysis of dual phase steels, Scripta Metall., 18 (1984) 327-332

[58] E. Navara, B. Bengtsson, K.E. Easterling, Austenite formation in manganese partitioning dual-phase steel, Mater. Sci. Technol., 2 (1986) 1196-1201

[59] Y. Toji, T. Yamashita, K. Nakajima, K. Okuda, H. Matsuda, K. Hasegawa, K. Seto, Effect of Mn partitioning during intercritical annealing on following $\gamma$-$\alpha$ transformation and resultant mechanical properties of cold-rolled dual phase steels, ISIJ Intl., 51 (2011) 818-825

[60] L. Barbé, Physical metallurgy of phosphorus alloyed TRIP steels, PhD Thesis, Ghent University, Belgium, 2005

[61] A.N. Kumar, S.N. Basu, Manganese partitioning and dual-phase characteristics in a microalloyed steel, J. Mater. Sci., 26 (1991) 2089-2092

[62] O. Dmitrieva, D. Ponge, G. Inden, J. Millan, P. Choi, J. Sietsma, D. Raabe, Chemical gradients across phase boundaries between martensite and austenite in steel studied by atom probe tomography and simulation, Acta Mater., 59 (2011) 364-374

[63] L. Cheng, C.M. Brakman, B.M. Korevaar, E.J. Mittemeijer, The tempering of iron-carbon martensite; dilatometric and calorimetric analysis, Metall. Trans. A, 19A (1988) 2415-2426

[64] G. Krauss, Martensite in steel: strength and structure, Mater. Sci. Eng. A, A273-275 (1999) 40-57

[65] F.D. Fischer, Q. Sun, K. Tanaka, Transformation-induced plasticity, Appl. Mech. Rev., 49 (1996) 317-364

[66] J. Chiang, B. Lawrence, J.D. Boyd, A.K. Pilkey, Effect of microstructure on retained austenite stability and work hardening of TRIP steels, Mater. Sci. Eng. A, 528 (2011) 4516-4521

[67] D.P. Koistinen, R.E. Marburger, A general equation prescribing the extent of the austenite-martensite transformation in pure iron-carbon alloys and plain carbon steels, Acta Metall., 7 (1959) 59-60

[68] S.M.C. van Bohemen, Bainite and martensite start temperature calculated with exponential carbon dependence, Mater. Sci. Technol., 28 (2012) 487-495

[69] S.M.C. van Bohemen, J. Sietsma, The kinetics of bainite and martensite formation in steels during cooling, Mater. Sci. Eng. A, 527 (2010) 6672-6676

[70] K.W. Andrews, Empirical formulae for the calculation of some transformation temperatures, J. Iron Steel Inst., 203 (1965) 721-729

[71] L. Barbé, M. De Meyer, B.C. De Cooman, Determination of the Ms$^\sigma$ temperature of dispersed phase TRIP-aided steels, in: International Conference on TRIP-aided High Strength Ferrous Alloys, Gent, Belgium, (2002)

[72] A.N. Vasilakos, J. Ohlert, K. Giasla, G.N. Haidemenopoulos, W. Bleck, Low-alloy TRIP steels a correlation between mechanical properties and the retained austenite stability, in: International Conference on TRIP-aided High Strength Ferrous Alloys, Gent, Belgium,

[73] P.J. Jacques, E. Girault, A. Mertens, B. Verlinden, J. Van Humbeeck, F. Delannay, The developments of cold-rolled TRIP-assisted multiphase steels: Al-alloyed TRIP-assisted multiphase steels, ISIJ Intl., 41 (2001) 1068-1074

[74] P.J. Jacques, E. Girault, P. Harlet, F. Delannay, The developments of cold-rolled TRIP-assisted multiphase steels: Low silicon TRIP-assisted multiphase steels, ISIJ Intl., 41 (2001) 1061-1067

[75] W. WANG, M. Zhu, Z. Cai, S. Luo, C. Ji, Micro-segregation behaviour of solute elements in the mushy zone of continuous casting wide-thick slab, Steel Research Int., 83 (2012) 1152-1162

[76] K. Mukhopadhyay, N.B. Ballal, N.N. Viswanathan, T. Venugoplan, P.K. Tripathi, P. Satya, Modelling of solidification & prediction of center line segregation in continuously cast steel slabs, Tata Search, (2012) 225-238

[77] ASTM ed., E1268-99 Standard practice for assessing the degree of banding or orientation of microstructures (reapproved 2001, 2007), in: ASTM ed. (Eds.), Annual book of ASTM standards (ASTM, 2000), Vol. 03.01 of metals-mechanical testing; elevated and low-temperature tests: metallography, 1999, pp. 820-848

[78] A. From, R. Sandström, Assessment of banding analysis in steels by using advanced image analysis, Mater. Charact., 41 (1998) 11-26. doi: 10.1016/S1044-5803(98)00019-9

[79] B. Krebs, A. Hazotte, L. Germain, M. Gouné, Quantitative analysis of banded structures in dual-phase steels, Image Anal. Stereol., 29 (2010) 85-90. doi: 10.5566/ias.v29.p85-90

[80] K.S. McGarrity, J. Sietsma, G. Jongbloed, Characterisation and quantification of microstructural banding in dual phase steels. Pt. 1: General 2D study, Mater. Sci. Technol., 28 (2012)

[81] K.S. McGarrity, J. Sietsma, G. Jongbloed, Characterisation and quantification of microstructural banding in dual phase steels. Pt. 2: Case study extending to 3D, Mater. Sci. Technol., 28 (2012)

[82] K.S. McGarrity, J. Sietsma, G. Jongbloed, Nonparametric inference in a stereological model with oriented cylinders applied to dual phase steel with microstructural banding, Ann. Appl. Stat., Submitted (2014)

[83] K.S. McGarrity, Stereological estimation of anisotropic microstructural features: Applying an oriented cylinder model to dual phase steel, PhD Thesis, Delft University of Technology, The Netherlands, 2013

[84] W.A. Backofen, Deformation processing, Addison-Wesley Publishing Company Inc., Reading, MA, USA, 1972

[85] E. Schmid, W. Boas, Plasticity of crystals, Hughes & Co., London, 1950

[86] J. Qu, W. Dabboussi, J. Nemes, S. Yue, Effect of grain size and martensite fraction on the dynamic factor of dual phase steel microstructures, Can. Metall. Quart., 48 (2009) 247-252

[87] C.A.N. Lanzillotto, F.B. Pickering, Structure-property relationships in DP steels, Met. Sci., 16 (1982) 371-382

[88] A.R. Marder, The structure-property relationships in Cr-bearing DP steels, in: R.A. Kot and B.L. Bramfitt (Eds.), Fundamentals of Dual-Phase Steels, Chicago, IL, USA, (1981)

[89] P.J. Jacques, On the macro- and micromechanics of TRIP-assisted multiphase steels, (2006)

[90] K. Sugimoto, M. Misu, M. Kobayashi, H. Shirawasa, Effects of second phase morphology on retained austenite morphology and tensile properties in a TRIP-aided DP steel sheet, ISIJ Intl., 33 (1993) 775-782

[91] A. Ramazani, A. Schwedt, A. Aretz, U. Prahl, W. Bleck, Characterization and modelling of failure initiation in DP steel, Comp. Mater. Sci., 75 (2013) 35-44

[92] H. Quade, U. Prahl, W. Bleck, Microstructure based hardening model for Transformation Induced Plasticity (TRIP) steels, Chem. Listy, 105 (2011) S705-S708

[93] A. Ramazani, K. Mukherjee, U. Prahl, W. Bleck, Modelling the effect of microstructural banding on the flow curve behaviour of dual-phase steels, Comp. Mater. Sci., 52 (2012) 46-54

[94] A. Ramazani, Z. Ebrahimi, U. Prahl, Study [of] the effect of martensite banding in the failure initiation in dual-phase steel, Comp. Mater. Sci., 87 (2014) 241-247

[95] J. Becker, E. Hornbogen, F. Wendl, Analysis of mechanical properties of a low-alloyed Mo-steel with different phosphorus additions, in: R.A. Kot and B.L. Bramfitt (Eds.), Fundamentals of Dual-Phase Steels, Chicago, IL, USA, (1981)

[96] I.B. Timokhina, P.D. Hodgson, E.V. Pereloma, Effect of microstructure on the stability of retained austenite in Transformation-Induced-Plasticity steels, Metall. Mater. Trans. A, 35A (2008) 2004-2331

[97] H.N. Han, C.G. Lee, C.S. Oh, T.H. Lee, S.J. Kim, A model for deformation behavior and mechanically induced martesitic transformation of metastable austenitic steel, Acta Mater., 52 (2004) 5203-5214

[98] Y. Tomota, H. Tokuda, Y. Adachi, M. Wakita, N. Minakawa, A. Moriai, Y. Morii, Tensile behavior of TRIP-aided multi-phase steels studied by in situ neutron diffraction, Acta Mater., 52 (2004) 5737-5745

[99] H.K.D.H. Bhadeshia, TRIP-assisted steels?, ISIJ Intl., 42 (2002) 1059-1060

[100] G.B. Olson, Effects of stress and deformation on martensite formation, in: Encyclopedia of Materials: Science and Technology, Elsevier Science, London, UK, 2001, pp. 2381-2384

[101] K. Sugimoto, N. Usui, M. Kobayashi, S. Hashimoto, Effects of volume fraction and stability of retained austenite on ductility of TRIP-aided DP steels, ISIJ Intl., 32 (1992) 1311-1318

[102] E. Jimenez-Melero, N.H. van Dijk, L. Zhao, J. Sietsma, S.E. Offerman, J.P. Wright, S. van der Zwaag, Characterization of individual retained austenite grains and their stability in low-alloyed TRIP steels, Acta Mater., 55 (2007) 6713-6723

[103] R. Blonde, E. Jimenez-Melero, L. Zhao, J.P. Wright, E. Bruck, S. van der Zwaag, N.H. van Dijk, High-energy X-ray diffraction study on the temperature-dependent mechanical stability of retained austenite in low-alloyed TRIP steels, Acta Mat., 60 (2012) 565-577

[104] E. Jimenez-Melero, N.H. van Dijk, L. Zhao, J. Sietsma, J.P. Wright, S. van der Zwaag, In situ synchrotron study on the interplay between martensite formation, texture evolution and load partitioning in low-alloyed TRIP steels, Mater. Sci. Eng. A, 528 (2011) 6407-6416

[105] Y. Takagi, R. Ueji, T. Mizuguchi, N. Tsuchida, Influence of strain rate on TRIP effect in SUS301L metastable austenite steel, Tetsu-to-Hagane, 97 (2011) 450-456. doi: http://dx.doi.org/10.2355/tetsutohagane.97.450

[106] H.Y. Yu, Microscopic response of TRIP steels to prestrain during plastic deformation, J. Iron Steel Res. Int., 20 (2013) 80-85

[107] A.H. Nakagawa, G. Thomas, Microstructure-mechanical property relationships of dual-phase steel wire, Metall. Trans. A, 16A (1985) 831-840

[108] J.P.M. Hoefnagels, C.C. Tasan, M. Pradelle, M.G.D. Geers, Brittle fracture-based experimental methodology for microstructural analysis, Appl. Mech. Mater., 13-14 (2008) 133-139

[109] C.C. Tasan, J.P.M. Hoefnagels, C.H.L.J. ten Horn, M.G.D. Geers, Experimental analysis of strain path dependent ductile damage mechanics and forming limits, Mech. Mater., 41 (2009) 1264-1276

[110] O. Leon-Garcia, R.H. Petrov, L.A.I. Kestens, Deformation and damage evolution of the microstructure around Ti particles in IF steel during tensile deformation, Key Eng. Mat., 348-349 (2007) 173-176

[111] H. Ghadbeigi, C. Pinna, S. Celotto, J.R. Yates, Local plastic strain evolution in a high strength dual-phase steel, Mater. Sci. Eng. A, 527 (2010) 5026-5032

[112] J. Moerman, P. Romano Triguero, C.C. Tasan, P. van Liempt, Evaluation of geometrically necessary dislocations density (GNDD) near phase boundaries in dual phase steels by means of EBSD, Mater. Sci. Forum, *702-703* (2012) 485-488

[113] J. Shi, S. Turteltaub, E. van der Giessen, Analysis of banded morphology in multiphase steels based on a discrete dislocation-transformation model, Modelling Simul. Mater. Sci. Eng., 19 (2011) 074006. doi: 10.1088/0965-0393/19/7/074006

[114] C.C. Tasan, J.P.M. Hoefnagels, M.G.D. Geers, Microstructural banding effects clarified through micrographic digital image correlation, Scripta Mater., 62 (2010) 835-838

[115] A.A. Korda, Y. Mutoh, Y. Miyashita, T. Sadasue, S.L. Mannan, In situ observation of fatigue crack retardation in banded ferrite-pearlite microstructure due to crack branching, Scripta Mater., 54 (2006) 1835-1840

[116] S.E. Offerman, N.H. van Dijk, M.T. Rekveldt, J. Sietsma, S. van der Zwaag, Ferrite/pearlite band formation in hot rolled medium carbon steel, Mater. Sci. Technol., 18 (2002) 297-303

[117] W. Xu, P.E.J. Rivera-Díaz-del-Castillo, S. van der Zwaag, Ferrite/pearlite band prevention in dual phase and TRIP steels: model development, ISIJ Intl., 45 (2005) 380-387

[118] M.G. Mecozzi, C. Bos, J. Sietsma, Microstructure modelling of solid-state transformations in low-alloy steel production, Mater. Sci. Forum, 706-709 (2012) 2782-2787

[119] C. Bos, M.G. Mecozzi, J. Sietsma, A microstructure model for recrystallisation and phase transformation during the dual-phase steel annealing cycle, Comp. Mater. Sci., 4 (2010) 692-699

[120] F.G. Caballero, A. García-Junceda, C. Capdevila, C. García de Andrés, Evolution of microstructural banding during the manufacturing process of dual phase steels, Mater. Trans., 47 (2006) 2269-2276. doi: 10.2320/matertrans.47.2269

[121] N.H. van Dijk, A.M. Butt, L. Zhao, J. Sietsma, S.E. Offerman, J.P. Wright, S. van der Zwaag, Thermal stability of retained austenite in TRIP steels studied by synchrotron X-ray diffraction during cooling, Acta Mater., 53 (2005) 5439-5447

[122] R. Blonde, E. Jimenez-Melero, L. Zhao, J.P. Wright, E. Bruck, S. van der Zwaag, N.H. van Dijk, High-energy X-ray diffraction study on the temperature-dependent mechanical stability of retained austenite in low-alloyed TRIP steels, Acta Mat., 60 (2012) 565-577

[123] M. Yonemura, T. Osuki, H. Terasaki, Y. Komizo, M. Sato, H. Toyokawa, A. Nozaki, Two-dimensional time-resolved x-ray diffraction study of dual phase rapid solidification in steels, J. Appl. Phys., 107 (2010) 013523-1-013523-6. doi: 10.1063/1.3277037

[124] J. Bischoff, A.T. Motta, R.J. Comstock, Evolution of the oxide structure of 9CrODS steel exposed to supercritical water, J. Nucl. Mater., 392 (2009) 272-279

[125] R. Blonde, E. Jimenez-Melero, L. Zhao, J.P. Wright, E. Bruck, S. van der Zwaag, N.H. van Dijk, High-energy X-ray diffraction study on the temperature-dependent mechanical stability of retained austenite in low-alloyed TRIP steels, Acta Mat., 60 (2012) 565-577

[126] D. Terada, M. Wadamori, N. Tsuji, Local deformation analysis in low carbon dual phase steel composed of martensite and ferrite, in: International Symposium on New developments in Advanced High-Strength Sheet steel, Vail, CO, USA, (2013)

[127] L. Salvo, P. Cloetens, E. Maire, S. Zabler, J.J. Blandin, J.Y. Buffiere, W. Ludwig, E. Boller, D. Bellet, C. Josserond, X-ray micro-tomography an attractive characterisation technique in materials science, Nucl. Instrum. Meth. B, 200 (2003) 273-286